# Gamma-ray bursts, BL Lacs, Supernovae, and Interacting Galaxies

Anup Rej
*Department of Physics (Section Lade), NTNU*
*7034 Trondheim, Norway*
Anup.Rej@phys.ntnu.no

## Abstract

*Within the framework of star formation in starburst galaxies undergoing interactions, connections among the red quasars, the BL Lacs, and the Blazars with the gamma-ray bursts are discussed in the light of the "hypernovae" scenario associated with the core-collapse of very massive Wolf-Rayet stars. Explanations for the occurrence of the gamma-ray bursts observed away from the center of the "host galaxies" are also provided. It is proposed that the gamma-ray bursts occur primarily in the star formation regions in starburst environments, and arise from the core-collapse of super-massive Wolf-Rayet stars, that are formed in interacting systems. The galaxies are formed mainly in two epochs corresponding to the redshifts $z \sim 3$ and $z \sim 1$. At the early epoch they are formed while the smaller systems interact and merge together to form larger galaxies. In this interacting environment star formations cause supernovae explosions. As the stars explode, shock waves propagate outward and collide with the ambient medium, forming a high-density super-shell, where intense star formation begins; subsequently supernovae explosions within the shell cause an outward expansion of the shell. Another scenario of star formation in relation to the gamma-ray bursts could be the interactions of large gas-rich low-surface brightness (LSB) spirals. The interactions among these galaxies give rise to fragmentation of their tidal tails. These fragmented clouds collapse to compact dwarfs, which undergo rapid star formation. Thus formed near the major mergers, such dwarf galaxies would fade away rapidly after the initial star bursts, and be found in the vicinity of the merged spirals. In such a scenario of interactions, where most quasars are observed, we discuss the MgII absorption lines in GRB 970508, whose spectra resemble a BL Lac object, and the spectral evolution of GRB 971214 that may also indicate a similar connection. Beside the BL Lac connection, the recent discovery of GRB 980425 in a nearby galaxy indicates asymmetric supernovae explosions that give rise to relativistic jets. We propose that the gamma-ray light curves arise due to inverse Compton scattering of soft-photons from precessing relativistic blobs of plasma moving in jets. This brings into light the mechanisms of supernova explosions, that give birth to relativistic jets*

# I. Introduction

Gamma-ray bursts (GRB's), believed at first to be a galactic phenomenon, remained an astrophysical enigma until the announcement (on February 28, 1997) of an optical counterpart of the GRB 970228 established their extragalactic origin. Further light was shed on this issue by the discovery of the counterpart of GRB 970508 in May 1997 which revealed strong metal line features at redshift $z=0.83$ and a weaker Mg II absorption system at $z= 0.768$, which bore great resemblance to the spectrum of the BL Lac object PKS 0138-097. The observation, in December 1997, of the counterpart of GRB 971214 brought this connection with BL Lac blazars even closer. The optical spectra did not show any strong emission lines during the gamma flare; such features emerged only after about two weeks. The blazars, on the other hand, show no emission lines during the outburst, but their spectra contain emission lines, like QSO's, when the flares subside. Furthermore only blazars, that are radio-loud and have highly polarized optical emission and flat radio spectra, emit high-energy gamma-rays. A new dimension to these phenomena was added with the discovery of SN 1998bw as a «bizarre supernova», which was associated with GRB 980425, discovered on April 15, 1998. Its optical spectrum could not be classified under the known classification scheme of the supernovae. The early spectrum did not resemble a SN Type Ib, or SN Type Ic. However the spectrum resembled a SN Type Ic at a later phase. It was found offset from the centre of a face-on barred spiral galaxy and a bright radio source was found coincident with it. The radio scintillation from the source indicated the presence of a relativistic jet. Apart from this connection between a supernova and a GRB, already in July 1992 a GRB was discovered two sigma from the supernova SN 1992ar. Within 3 sigma error box with GRB 971115 and GRB 971120 a supernova SN 1997ef was found on November 25, 1997. The most luminous supernova, SN 1997cy was discovered less than a degree away from the GRB 970514. In fact, Wang and Wheeler (1998) reported a strong correlation between SN Type Ib supernovae and GRB's and suggested that SN Type Ib/Ic should be important sources for gamma-ray bursts.

An earlier paper (Rej & Østgaard 1997) suggested that the gamma-ray bursts result from «failed Type I supernova/hypernova» in high redshift starburst galaxies, and proposed that the two peaks observed in the BATSE data of GRB's are associated with the two peaks of star formations that are observed at redshifts $z \sim 3$ and $z \sim 1$. The present article extends the previous analysis to the observations that have since been reported, and proposes connections among gamma-ray bursts with BL Lacs and red quasars, that can be grouped under blazars. It discusses the features of quasars, BL Lacs and blazars, that arise in the environments of interacting galaxies at high redshift, and relates various mechanisms by which the Mg II absorbing systems may be formed in the environment of interactions as the tails of low surface-brightness spiral galaxies are stripped off. The fragments from such disruptions may generate compact regions of star formations that rapidly fade with Type Ib or Type II supernovae explosions. The analysis is based on a scenario that includes dwarf systems which give birth to rapid starbursts in the vicinity of low surface brightness galaxies, and fade away by explosions of very massive Wolf-Rayet stars. Such explosions may generate relativistic jets that are beamed towards us. Inverse Compton scattering of soft photons, emitted by the surrounding stars, and/or by the jet itself, and/or from the centre of the collapsed system, may give rise to the observed gamma-ray light curves as flares from blobs of plasma that rotate as magnetized knots move within the jet. In the Lorentz boosted situations they will appear as intense gamma-ray bursts.

The rest of the paper is organized as follows. Section II reviews the observational data of the gamma-ray bursts whose counterparts are detected. The next section relates the properties of quasars, BL Lacs and blazars with a view to elucidating the connection between the GRB's and the objects that are grouped under the category called blazars. The mechanisms of gamma-ray emissions from relativistic jets and the correlation among radio, optical, X-ray and gamma-ray emission arising from synchrotron self-Compton process are discussed in section IV. An explanation of the gamma-ray burst profile arising from inverse Compton scattering of soft photons from blobs moving in precessing jets will be given. The reason for the long-lasting X-ray afterglow after the blast are also investigated. Sections V and VI treat the relations between quasars and star formation in starburst galaxies, at local as well as high-redshift universe. The possibility of rapid formation of very massive stars and their subsequent explosions in dwarf systems, that harbour massive low surface brightness galaxies, is also explored. Such explosions in the starburst galaxies are shown to be related with the MgII absorbing clouds, that are injected into the Inter Galactic Medium (IGM). The concluding section forges an understanding between the observed data about the counterparts and the epochs of galaxy formation in the universe, and point to the necessity of investigating the mechanism of explosions of very massive stars that may eject relativistic jets.

## II. The optical counterparts

### II.1. GRB 970228

The optical counterpart of GRB 970228 (Groot et al. 1997, Galama et al. 1997, Van Paradijs et al. 1997), the first ever detected, showed a decaying «fireball», that plummeted from 21st to 23rd magnitude in eight days. Afterwards, it could not be detected by ground-based telescopes. The bulk optical event ( in the rising and the first decay phase) developed in more than 3d.6. The maximum of the «fireball» emission was attained probably not earlier than 0d.71 after the gamma-ray burst. If one assumes a power-law for the luminosity decay, i.e., $L_{opt} \propto t^{-\alpha}$ a limit $\alpha > 1.1$ can be determined from the HST data of March 26. However, the optical data of the first days require $\alpha > 1.4$.

The Hubble observations on March 26, 1997 (Sahu et al. 1997) showed that the visible GRB source has a point-like structure together with an extended one. The colour indices reddened significantly during March ( Guarnieri et al. 1997). At its maximum, the *R*-luminosity of the optical transient was about 15 times that of the underlying extended object. Between March 1997 and September 4, 1977, the brightmess of the visible poin-like component diminished by a factor of 500. The point source is still fading according to a $t^{-1}$ power-law at a fixed position. The Hubble telescope continues to see the «fireball» and a surrounding nebulosity (at 25 mag) which is considered to be a dwarf host galaxy at a redshift ~ 1. The continued visibility of the burst and the rate of its fading support the theory that the light from the burst is an expanding relativistic «fireball». The proper motion observed in the optical counterpart could be interpreted as due to the motion of a relativistic plasma ejected in a jet-like geometry ( Caraveo et al. 1997). Moreover, the HST observations clearly show that the optical counterpart is not located at the centre of the brightness distribution, suggesting that it is not related to the AGN activities at the centre of the host galaxy. A close examination of the host galaxy reveals elongated features to the east, which are suggestive of a spiral rather than an elliptical structure. Furthermore, the *V-I* colour of the extended source suggests a late-type galaxy.

Although the rate of decay of the visible component of the supernovae spectrum is comparatively slow, that of the UV component amounts to 2.5 mag/day. Thus, a supernova at the appropriate redshift ($z$ ~ 1-2) could appear to decay in the visual (UV rest frame) at the observed rate. The observed fast decline would imply a redshift $z > 1$ of this supernova. Such high redshifts would make this object brighter than the brightest supernovae observed either at low or at high redshift by about two magnitudes at the peak.

The BeppoSAX narrow-field X-ray telescope was pointed to the burst after 8 hours of its discovery ( Costa et al. 1997b). The X-ray flux was $(2.8\pm0.4) \times 10^{-12}$ erg cm$^{-2}$ s$^{-1}$ in the 2–10 keV band and $(4.0\pm0.6) \times 10^{-12}$ erg cm$^{-2}$ s$^{-1}$ in the 0.5–10 keV band. A second follow up was done on the X-ray source (ISAX J0501.7+1146), at the position R.A. 5h01m44s, decl. +11° 46'.7, after two days. The second observation showed twenty times lower source flux compared to the first . A decay index $\alpha \sim 1.4$ is found. The ratio L opt(R)/Lx(0.5-10keV) $\approx 4 \times 10^{-3}$ (if the luminosity is taken at the observed maximum , at least 13 hours apart). By scaling the X-ray flux at the time of the optical maximum, one obtains a ratio $\approx 6 \times 10^{-3}$. The upper limit of the ratio is set at $\approx 1.2 \times 10^{-2}$.

The light curve displays a strong and rapidly variable peak about 4 s long, followed by a much weaker and smoother peak which begins 40 sec later ( Costa et al. 1997a) The first pulse is variable on a time scale of 0.1 sec and may have been produced by the shock, while the second pulse signals the beginning of an afterglow produced by the external shock. The overall duration of the first pulse corresponds to the duration of intense activity of the "inner engine" which emitted the ejecta. The variability in time scale of 0.1 sec sets an upper bound on its size of $3 \times 10^9$ cm.

The X-ray (2-30 keV) light curve displays two comparable peaks, corresponding in time to the gamma-ray peaks. X-ray emission, soon after the GRB, continuously evolves into the X-ray emission afterglow. The observations made after 8 hours showed that the source energy spectrum in 0.1-10 keV is consistent with a power-law of photon index 2.1 ±0.3. After three days the 2-10 keV flux was $(1.5\pm0.5) \times 10^{-13}$ erg cm$^{-2}$ s$^{-1}$. The upper limit in the 0.1-2 keV band was $4 \times 10^{-13}$ erg cm$^{-2}$ s$^{-1}$.

The GRB integrated flux measured in 40-700 keV was 1.1 x $10^{-5}$ erg $cm^{-2}$. The X-ray fluence measured in 2-10 keV was about 1.2 x $10^{-6}$ erg cm-2. The X-ray afterglow is not only the low energy tail of the GRB phenomenon, but it is a significant channel of energy dissipation of the event on a different scale.

## II. 2. GRB 970508

*Radio and millimeter observation*
D. Frail ( Frail et al. 1997) of the National Radio Astronomical Observatory discovered the first radio emission associated with any GRB at 8.46 GHz with VLA observation. Also VLBI observation was made by Taylor (1997) at 8.41 GHz. Pooley and Green (IAUC 6670) detected the source at the MRAO at 15 GHz on May 16.5 and on May 17-22 . Gurendl et al. (1997) observed the source at 86 GHz with the BIMA and Shepherd et al. (1997) used OVRO to detect radio emission at 86.8 GHz.

The radio data of Frail show approximately constant average fluxes with many superimposed flares. Thirty days after the burst these flares have calmed down. The variation could be either intrinsic or extrinsic. The intrinsic case demands shock induced coherent/collective plasma emission as in intra-day variable QSO's (Standke et al 1996). The emission spectrum is $\alpha \propto \nu^{1.1}$. This spectrum has a steep slope and can be interpreted as the transition between optically thick and thin conditions in the frequency range observed. There is a change in the scintillation pattern of the radio emission, which confirms a relativistic expansion.

Bremer et al. (1998) detected between 17-22 May a continuum point source ( 1.62 ± 0.25 mJy) at 86.2 GHz with IRAM, and found no detection at 232 GHz. The source seemed to have passed from detection to non-detection in the millimeter range between days 14 and 19 after the burst, without showing correlated flares in the optical, or cm range. They observed significant fading which is not well fitted with the $t^{-1.1\pm0.1}$ power law that is found at higher frequencies.

*Optical observation*
Initial optical images were obtained only 5.8 hours after the burst was detected ( Bond 1997 and Djorgovski et al. 1997b). The optical source brightened from May 9 to May 10, reaching r ≈ 20 mag and then started declining. The optical magnitude between 1.3 and 1.9 days after the burst suggested that at this time the characteristic synchrotron frequency of the afterglow passed through the visible spectrum. The optical afterglow was a few hundred times more luminous than the brightest supernova. A few months after the peak, it still remains more luminous than any supernova. Assuming spherical emission, its total energy was ~ $10^{52}$ ergs. The decline in the brightness can well be fitted with a $t^{-1.14}$ power-law. The total integrated flux in the red light is $I_r$ >4.6x$10^{-8}$ erg cm-2. The optical spectral index is $\alpha$ = -0.9± 0.3.

The Hubble observation, made on June 2, 1997 ( Pian et al. 1997), showed no accompanying object such as a host galaxy near the source. However, a few faint galaxies several arcseconds from the source were seen. A faint blue galaxy with g ≈ 24.5± 0.4 mag is clearly detected at 4.3 arcsec east and 3.5 arcsec north of the optical counterpart (Djorgovski et al. 1997a). The magnitude of this galaxy is typical of the general field galaxy population at z ≈ 0.8, and its blue colour is suggestive of active star formation. If it is the host, then the burst is well outside the galaxy's stellar disk.

The counterpart was monitored regularly in $R_c$ until about 4 months after the burst. The optical spectrum became redder before reaching maximum light. A magnitude $R_c$=26.09±0.39 was also suggested for the possible underlying galaxy. The optical fluence (3000-10000Å) is about 5% of the BATSE and 20% of the WFC fluence. A comparison with X-ray afterglow ( Piro et al. 1997a) $S_X$=7.3 x$10^{-7}$ erg cm-2 2-10 keV showed that the optical afterglow fluence is about 20% of the X-ray afterglow fluence.

The spectroscopic observation of the counterpart (OT J065349+79163) shows mostly a featureless spectrum except for a few prominent absorption lines. Absorption line features suggest relatively strong metal line absorption features at z=0.835 and a weaker Mg II system at z=0.768 (Metzger et al. 1997a,b,c). There is absence of a Ly$\alpha$ forest, which implies that the source lies at 0.835≤ z≤ 2.3. The presence of Mg II and Fe II absorption lines and absence of detectable C IV in the spectrum of the afterglow limit the redshift of GRB 970508 to z< 1.8.

The continuum source is either more distant and absorbed by a gas cloud at this redshift, or is, perhaps, located within the cloud. Such absorption systems are commonly seen in the spectra of high-redshift quasars, which show either Mg II (singly ionized magnesium), or C IV (triply ionized carbon) absorption. Most of the absorption are associated with normal galaxies close to the line of sight of the QSO's. At these redshifts, the number of Mg II absorption systems are usual. However, the ratio of the strength of Mg I/Mg II in this case appears unusually high. Combined with the high strength of the Mg II, the absorption system provides an evidence for a dense foreground interstellar medium.

Moreover, the [O II] 372.8 nm emission indicates a normal interstellar medium rather than AGN's. The emission line region is also very compact. Therefore the probability that the positions of the optical region and the line emission regions coincide by chance is small. It is reasonable to believe that the two are related, and z=0.835 could be the GRB's redshift and not just a lower limit.

The compactness of the region makes it a good candidate for a star-forming scenario. Otherwise, the source's compact optical appearance, featureless continuum, X-ray emission, and high redshift, suggest a possible association with BL Lacs (Stocke et al. 1991).

*X-ray and gamma-ray observation*
The burst was first detected by BeppoSAX (Costa et al. 1997c, Heiese et al. 1997a). The BATSE 20-1000 keV fluence was $(3.1 \pm 0.2) \times 10^{-6}$ erg cm$^{-2}$ and a peak flux density (50-300 keV) was $(1.66 \pm 0.06) \times 10^{-7}$ erg cm$^{-2}$ s$^{-1}$ (Kouveliotou et al. 1997). It had an integrated flux $I_\gamma \approx (1.8 \pm 0.3) \times 10^{-6}$ erg cm$^{-2}$ in the energy band 40-700 keV.

The X-ray maximum occurred about $10^3$ s after the burst (before the BeppoSAX follow-up observation). However, the X-ray emission measured simultaneously with the GRB was, probably, the low-frequency extrapolation of the gamma-rays produced by the internal shocks, and was not the beginning of the afterglow. The fluence recorded with WFC (2-26 keV) was $(0.7 \pm 0.1) \times 10^{-6}$ erg cm$^{-2}$ (Jager et al. 1997, Piro et al. 1997b).

After the initial burst the X-ray emission decayed following the power-law $\sim t^{-1.4}$ upto the time $\sim 6 \times 10^4$ sec (Piro et al. 1998). It was then followed by a burst with a duration of $\sim 10^5$ s with a time behaviour similar to what was observed in the optical.

## II.3. GRB 971214

It was a structured burst (Heiese et al. 1997b) lasting about 25 sec with a peak flux of 650 counts/s. The peak of the X-ray flux was about 1 Crab. A follow-up observation, nearly 7 hrs after the burst, showed a X-ray source (ISAXJ1156.4+6513) at that position, which faded by a factor of five in several hours.

*Radio and millimeter observation*
Frail and Kulkarni et al. made observations at a frequency of 8.46 GHz about a day after the burst and found no radio emission.

*Optical Observation*
Optical observations were made by Halpern et al. (1997), Castander et al. (1997) and Rhoads (1997). I-band images, revealing the counterpart, were obtained by John Thorstensen 12 hours and 36 hours after the burst. It revealed a point-like transient, that faded from approximately I=21.1 on the first night to near the magnitude of I=22.6 on the second night. On December 17, 1997 Caltech GRB collaboration obtained spectra of the optical transient and reported the presence of no strong emission-lines from the transient. However, the spectra of December 28, 1997 showed a slightly extended emission feature at 5384 A. Additionally a broad absorption feature was seen at 5752 A.

APO observation in R-band showed brightness to 24.6 mag after 3.7 days (December 18). The brightness was measured to be 22.06 after 0.71 day. Caltech group conducted observations on January 10, 1998 in the R-band and the magnitude of the host galaxy was predicted to be 25.6 mag. This is ~ 2 mag brighter than the extrapolated magnitude of the afterglow. By identifying the emission feature as Lyman-alpha, and the absorption feature as OI 1302 A, the host galaxy was found to be at redshift 3.43.

Diercks et al. (1997) reported R-band light curve measurements, made 17 hours after the burst, showing a power-law decay with index $\alpha = -1.38$. It was significantly steeper than the decay of the optical counterparts of GRB 970228 ($\alpha=-1.1$) and GRB 970508 ($\alpha=-1.14$). The brightness of GRB 971214 was also ~ 1 mag fainter than GRB 970228. They noted that there was a possibility that the colour of the afterglow may change with time and the time dilation may move various features of the lightcurve.

*X-ray observation*
The BATSE event consisted of a complex series of pulses lasting about 40 sec, with an estimated total fluence above 20 keV of $1.09(+/- 0.07) \times 10^{-5}$ erg cm$^{-2}$ and a peak flux occurring at trigger 13 sec of $1.95(+/-0.05)$ photons cm$^{-2}$ s$^{-1}$ (50-300 keV; 1.024-s integration).

The event was also observed by a single RXTE-ASM camera, which detected a flux enhancement lasting 50 sec and reaching a peak intensity of 470 ± 140 mCrab ( 1.5-12 keV; 1-s integration), 15 s after the burst trigger. The gamma burst was triggered a few seconds earlier.

## II. 4. GRB 980326

GRB 980326 was detected (Celidonio et al. 1998) on March 26.888 UT with WFC, GRBM and BATSE (Briggs et al. 1998). In the BATSE energy range (25 -1800 keV) the event lasted about 5 sec and was resolved into three narrow peaks with a peak flux of $8.8 \times 10^{-7}$ ergs cm$^{-2}$ s$^{-1}$ over a 1s time scale. Its total keV fluence was $1.4 \times 10^{-6}$ erg cm$^{-2}$. Its $E_{peak}$ was unusually low ( 47 ± 5 keV). Only 4% of bursts have similar low values.

Optical $R_c$ band observation started, among others, at AAT on march 27.40 UT, followed by observations at the NTT (New Technology Telescope) and Danish Telescope at ESO. R-band image taken using Keck-II 10-m telescope on the night of 17th April 1998 detected a galaxy coincident within less than 0.3 arcsec with the OT reported by Kulkarni et al.(1998). It exhibited a temporal decay, which could be fitted with a power-law and a constant source. The power-law index is $\alpha$=2.1 ± 0.13, which is by far higher than that of the previous afterglow. The light curve exhibits flattening which is possibly a signature of an underlying galaxy with magnitude R=25.5 +/- 0.5. Grossan et al. (1998) reported an elongation in the NE-SW direction. However the observations reported by Djorgovski et al. (1998) displayed an elongation in exactly the perpendicular direction (SE-NW) which may be due to fading of the OT.

RXTE/PCA scanning ( Marshall and Takeshima 1998), 8.5 hours after the burst, sets the upper limit of $1.6 \times 10^{-12}$ erg cm$^{-2}$ s$^{-1}$ on the 2-10 keV X-ray afterglow.

## II. 5. GRB 980425

Soffitta et al.(1998) reported the detection by BeppoSAX monitor of a gamma-ray burst on April 25.90915UT. The BATSE confirmed the detection (Kippen et al. 1998). In terms of BATSE burst/flux distribution the event was an average type. BATSE saw the burst (Galama et al. 1998a) for about 35 sec, and inferred a total energy of $(8.5 \pm 1.0) \times 10^{47}$ erg in gamma-rays (or $10^3$-$10^4$ times fainter than a typical GRB). It saw no emission above 300 keV. At this luminosity other GRB's would have been invisible had they occurred at a distance 20 times farther away.

On April 28, the flux density of the source were 9 and 13 mJy at 6 and 3 cm, respectively, and were similar on April 29 (9.9 and 13 mJy). On May 5 the flux densities of the source increased dramatically to 39 and 48 mJy at 6 and 3 cm, respectively. The position of the variable radio source coincided with the optical position given by Galama. At the distance to ESO 184-G82 the radio source is already three times more luminous than SN 1988z, one of the most luminous radio supernova discovered, and is still brightening. It is believed that the radio source is related to GRB 980425.

No X-ray emission was observed. Kulkarni et al. (IAUC 6899) suggested that the radio emission may arise in a relativistic shock, while the optical emission may come from low velocity shock.

Within the localization of GRB 980425, Galama et al. (1998b) reported a supernova, possibly of Type Ib (IAUC 6901). The object appeared to have exploded on, or around April 24, 1998.

When first observed, 0.6 days after the GRB, the supernova was falling in brightness. But afterwards, it started becoming brighter. It has continued to brighten as follows: May 8.311 UT, V=13.87; at May 8.306 R=13.84 and at May 8.309 I=13.98. The optical spectra resemble those of SN 1997ef, which can not be classified under the current classification scheme of supernovae. Woosley et al. (astro-ph/9806299) found that the supernova is well modeled as the explosion of a carbon-oxygen core of 6 solar mass. They point out that, when viewed at certain angles, massive stars and unusual explosions could only make such a GRB. In their opinion most GRB's are probably beamed phenomena, while supernovae are visible at all angles.

## II. 6. GRB 980519

The optical transient was first discovered by Jaunsen et al. (1998). Djorgvski et al. (1998) obtained multicolour CCD images using Palomar Observatory 200-inch telescopes on May 20. The object is continuing to fade.

The R-band data of Gal et al.(1998) using Keck Observatory 10-m telescope on May 21 show that OT has R=23.48+/- 0.2. The OT host galaxy magnitude is not brighter than R ~ 24. The R band magnitude is consistent with the source still fading with a slope of approximately 2.0.

## III. Mg II systems, Quasars, BL Lac Blazars

### III. 1. Quasars

*Optical Transient (OT) and QSO*
The first indication of a relation between QSO's and GRB's came with *GB 910219,* which was detected by WATCH aboard the Soviet Space Observatory GRANAT . It consisted of two main bright peaks, separated by about 50 s with a duration of 10 s each. The estimated fluence was 10-5 ergs cm-2 between 6 and 120 keV.

The error box of localization of GB 910219 was related to the OT error ellipse that included a QSO with z=1.78.

*Characteristics*
The standard definition of a quasar involves the following prime characteristics:
1) A stellar appearance
2) A strong ultraviolet excess
3) A variable optical emission
4) The broad (>20° in QSO rest frame) permitted emission lines. High redshift QSO's also exhibit absorption line features
5) Large redshift and absolute magnitude -23 to -28, which is much brighter than even the brightest cluster galaxies ( -22 to -24).

The bright quasars exhibit remarkable permanence as to the spectra. They are similar for QSO's at redshifts in the range z≈ 0-5.

Quasars show apparent separation velocities between 3 and 45 times the speed of light. General consensus asserts that either ejected blobs of plasma are moving at relativistic velocities, or there are jets whose contents are flowing out at relativistic speeds.

*Morphological types of the quasar host galaxies*
The host galaxies are centred on the quasar to the accuracy of measurements (~400 pc). There are more radio-quiet quasars embedded in ellipticals than in normal spirals. A sample of quasars with 0.44<z< 0.828  reveals that QSR's appear to be embedded in giant ellipticals, with average absolute magnitude of -23 and an average diameter of 130 kpc. The luminous quasars occur preferentially in luminous galaxies. The average absolute magnitude of the hosts is 2.2 mag brighter than expected for a field galaxy. QSR's are 1.7 mag brighter than QSO's and are located in galaxies 1.5 mag brighter. Radio loud quasars also possess a slightly larger redshifts. At least 25% of quasars are found in compact groups or clusters of galaxies.

Observations of luminous quasars, at redshifts smaller than 0.3, show(Bahcall et al. 1997) that  luminous quasars occur in diverse environments: In ellipticals as bright as the brightest cluster galaxies, in apparently normal ellipticals and spirals, and in complex systems of gravitationally interacting components and faint surrounding nebulosity. Over 40% show some spiral characteristics, with the rest being of undetermined morphology.

*Quasar environments*
out to z ~ 0.3 upto half of the host galaxies are either found to be engaged in galaxy interactions and merging events, or possess close neighbours. At least 30% are seen to be interacting with the nearby galaxies. At somewhat wider range, the environments are found to comprise 10-20 galaxies on the average (Fisher et al. 1996).

*Radio sources*

Their radio output far exceeds the normal galaxies. However the majority of quasars are radio quiet. As the radio quiet quasars are more difficult to discover, most quasars are found to be radio emitting type. Many also strongly emit X-rays and are believed to be Seyferts having extremely luminous nuclei, that obscure all other stars in the galaxies, thus rendering them quasi-stellar appearance.

The central radio emitting regions are similar to other AGN's. These compact radio sources have small angular size (~0.1'' to ~0.0001'', that is the limit of the VLBI resolution), high surface brightness with peak brightness temperatures upto ~ $10^{12}$ K, significant variability over months to years, and flat, undulating or inverted spectra between ~10 MHz and ~100 GHz. ( In contrast most extended sources show either a fairly constant spectral index of $\alpha \approx -0.8$, or one that steepens at higher frequencies).

Radio emissions emerge from somewhat extended regions as well as from the compact cores. There exist core-jet structures in the central parts of the QSR's: The jet of low surface brightness, with moderately steep spectra, are capped by bright spots with flatter spectra. However, the cores are more prominent at higher frequencies. Because of the large difference in brightness between the core and the jet it is hard to map the jets. In 3C 345, where the mapping has been made, the jet appears to be broken into bright patches.

Compact cores are detected in 35% of the radio sources, and 80% of the samples of quasars exhibit cores that are at least partially resolved. The percentage of radio flux in the core increases with the radio spectra index, as the radio variability increases. Quasars, with extended one-sided emission on a larger scale, appear to have a significantly greater fraction of the flux in the cores, than do normal double QSO's. The cores of double QSO's show an inverse correlation with the linear size of the double structure. The flux from the core shows correlation both with the apparent misalignment of the hot spots in the lobes from co-linearity through the central component, as well as with the ratio of the separations of the hot-spots from the centre. These are in accord with the beaming models for the compact sources.

1 mm emission is observed from all radio-loud QSR's (but not from the radio-quiet ones). The mm fluxes are consistent with a direct extrapolation of the radio continuum. The radio properties of the quasars are correlated with their optical spectra.

The redshift distribution of 1 Jy quasar sample peaks at z ~ 1 while the quasar samples derived from radio surveys with lower flux density limits peak close to the redshift peak which is found for the optically selected quasars.

The fraction of the optically selected quasars, which are flat-spectrum radio sources, appears to be roughly constant with redshifts at z>1.

*Variability*
Some sources show individual or multiple outbursts that boost output by a factor of two or three for a year or so, while others may take a decade, or more, to vary through the same factor. These variations are seen simultaneously throughout the radio band, and very often the strength of the outburst is quite independent of the wavelength. Flux variations are pronounced in most compact sources. Such behaviour may be expected from optically thin synchrotron sources, where the flux increase is due to continuous ejection and /or acceleration of relativistic particles, and the decrease is due to adiabatic, or radiation losses. The flux rises first at higher frequencies, and then increase by lesser amounts towards the lower frequencies. This can be explained by an expanding source of synchrotron emission that starts out opaque and becomes transparent at lower frequencies. Another type of flare has been seen (in 3C273), where an outburst, detected both in mm and IR, liberated around 0.1 $M_\odot$ $c^2$ during the course of two months. The spectral index of the flare seemed to be constant $\alpha \approx -1.0$. The change in luminosity of a factor of two within one day (seen at 1 mm in 3C 273), demand relativistic bulk motion near the line-of-sight, and implies that it should be classified as an optically violently variable (OVV) type.

However there is no significant difference in optical variability between radio quiet and radio loud ones. One significant difference is that the QSR's are more powerful X-ray emitters.

*Optical*
Roughly 10% of radio-loud quasars are OVV. They can change their optical fluxes by a factor of 2 in less than a week. OVV quasars can display high ( 10%) linear polarization, which, on occasion, can vary even more rapidly. Most other quasars show less than 1% polarization. Those low-polarization quasars, that also possess double lobes, exhibit a correlation between the optical polarization direction and the radio source axis.

Aside from the OVV quasars, polarization are usually small, and hard to detect. Among the brightest quasars about 90% have linear polarization < 2% , and much of this could be due to interstellar polarization in our own Galaxy. Radio-quiet quasars almost never exhibit strong polarization. This argues against the presence of much dust in them.

*Infrared*
Typical QSO energy output peak in the IR, or mm domain. The total IR-optical powers range between ~ $10^{11}$ and ~ 3 x $10^{11}$ $L_\odot$ . In IR the similarities among the quasars are strongest, with over-all power law slopes of - 1 to -2. However, there is some flattening towards the far IR.

Several low redshift QSO's (including 3C 273) show a Balmer recombination component. The high redshift QSO's have their Lyman continuum edge shifted into the observable range. Many of them show continuum absorption too.

*UV*
In near UV, most QSO's seem to depart. The "blue bump" is a broad emission feature between 2500 and 4000 Å. The addition of a Balmer continuum to a background nonthermal power-law can explain this bump. In high-redshift UV radiation can be detected in the optical domain.

**Emission lines**

In low redshift cases, both broad emission lines, equivalent to Doppler broadening of > 5000 km s$^{-1}$ , and narrow emission lines due to forbidden transitions with FWHMs between 300 and 1500 km s$^{-1}$ are observed.

*Broad emission line region*
In the broad line emission regions (BLR) the operative processes are photoionization and heating of clouds within the central parsec. The photons come from a UV as well as soft X-ray continuum produced at the core. Such photoionizing continuum has a power-law. The size of the BLR is roughly proportional to the square root of the luminosity, that is capable of ionizing the gas (as for Stromgren sphere). It corresponds to less than, or about, 1 pc. The high-order Balmer lines, the Balmer continuum and the numerous FeII lines form an apparent continuum, superimposed on the continuum due to the central source. Most models, explaining BLR, involve compact nonthermal region, partially surrounded by absorbing clouds. Assuming that the photoionization is dominant , BLR is optically very thick (only optically thick models can yield Mg II lines as observed). Since much of the continuum is visible, the blobs, that make up BLR, cover only a fraction of the surface of the innermost parsec. The total mass of the gas in BLR is estimated to be $10^2$ to $10^3$ $M_\odot$ . This divided among many small clouds (~$10^{-9}$ $M_\odot$), will give only a few percent covering factor. These cloudlets may have a wide spread in velocity, with some approaching us, and others receding from us, giving rise to huge line widths. The composition of the gas is basically the same as the sun, although carbon, nitrogen and oxygen may be more abundant. They are almost certainly confined by a hot intercloud medium, and perhaps by relativistic particles as well. The clouds within the BLR fall into outflow and gravitationally dominated models. The outflow consists of either ballistic outflow, or radiatively accelerated ejection. It requires gas to condense into clouds very close to the continuum source at the very centre. However such models require very high confining mass > $10^{10}$ $M_\odot$.

*Narrow emission line region*
The narrow emission lines are difficult to detect. They arise significantly outside the core. These lines are interpreted as originating over a volume out to 1 kpc. To explain these lines either the whole region contains fast turbulent motions, or any cloud within it must be moving very rapidly. Irregularities and asymmetries imply that NLR is inhomogeneous, and may have a net motion relative to the continuum source.

*Absorption lines*
Absorption lines are found in all high redshift QSO's. At least four classes of line systems exist. Type A's consist of very broad absorption troughs with z's different from $z_{em}$ by upto 0.1 c. Type B's are the ones with sharp lines having absorption redshifts not very different from the emission redshifts. Types C's and D's are those with sharp lines with $(z_{em} - z_{abs})/(1+z_{em}) > 0.01$. C-system contain metal lines, but the D-system consist of larger number of Lyman absorption lines seen at wavelengths below the Lyα emission. These types of absorption lines are due to clouds in the galaxies, especially in the halos, that are between the QSO's and us. The great majority of absorption lines arise outside the emission lines region. However many absorption line regions can be very close to QSO's. Perhaps they arise from clouds in, or near, galaxies in a cluster that also contains the QSO's.

*X-ray and gamma-ray properties*
Most radio-loud quasars are X-ray emitters, giving off $10^{43}$ to $10^{47}$ ergs s$^{-1}$, between 0.5 and 4.5 keV. $L_x/L_{opt}$ may range from 0.02 to 3. The average slope of the power-law, that fits between the optical and the X-ray bands, is -1.46 (although values between -1.0 and -1.7 are known). Power-law fit between 10 and 50 keV are available for 3C 273 and QSO 0241+622. The best values for the index are -.41 and -0.93 respectively. Between 1- 10 keV power-law with index -0.8 fits the data, and is consistent with indices observed for Seyfert galaxies and other AGN's. Since there may be a common power-law fit from IR to soft X-ray of index -1.1, the shallower spectral index for X-ray may indicate two distinct X-ray components in most QSO's.

Significant variability on time scales of day is rather common, but not more rapidly. Over time scales of 6-18 months a majority of QSO's show significant changes (though not variation of factor of two or three). A factor of three flux change within 200 secs has been observed in QSO 1525+227. In the same galaxy after 13 months no short term variability was detected.

### *Correlation between different wavelengths*

*Optical and radio*
Most optically selected quasars show no evidence for the synchrotron emission that produces the bulk of the radio output. Absence of strong high-frequency radio power in optically bright QSO's could be due to the size of the sources that are so compact as to be self-absorbed, or the radio could be attenuated by free-free absorption due to thermal gas in the BLR. Among the QSR's over 80% show compact milliarcsec cores, and the pieces of their total flux, coming from that core, rise for the optically redder quasars but do not seem to depend much upon the redshift. A single power-law rarely fits the entire optical continuum because of the massive line blending and a probable quasi-blackbody component. In these sources, cm-spectra are flatter, implying a break in the spectrum at < 1mm.

*X-ray and mm*
Interesting correlation exists between X-ray and mm flux. It could be due to synchrotron self-Compton radiation. X-ray could come from mm photons scattered off relativistic electrons. However, some sources vary so rapidly in mm that any such correlation should be broken. Perhaps a synchrotron jet sources may be the cause!

*X-ray and radio*
Correlation exists between radio and X-ray powers. For a given optical luminosity, the X-ray luminosity for the radio-loud quasars are about three times more than in the radio quiet ones. The radio and X-ray loud quasars are more variable. In the radio quiet QSO's X-ray luminosity may come from a very compact region, which is optically thick in the radio, but the QSR's have jets in addition, or synchrotron self-Compton components, which add to X-ray emission.

### *Red quasars and BL Lacs*

About 10% of the flat-spectrum radio quasars are red quasars. Selecting the reddest of the sample one finds that about one quarter of them are BL Lacs. The discovery of molecular line (Carilli et al. 1998) emission from a sample of red quasars indicates the presence of dust, possibly concentrated in dense nuclear tori.

The red quasars have faint, extended, galaxy-like emission in the optical, with little or no evidence for bright point source, while near IR images reveal a dominant point source in most cases. Most of them are variable in near IR.

The red quasar 0108+388 is associated with a narrow emission line galaxy at z=0.67 with R=22.0 mag. The optical image shows a very red, diffuse and slightly asymmetric galaxy, perhaps a face-on spiral (Carilli et al. 1997). This GHz peaked radio source is a compact symmetric object, with an inverted spectrum in the nucleus, plus twin jets with steeper spectra extending 3 mas from the nucleus towards the northeast and southwest. The radio source projects within 0.5" of the centre. However, there is no evidence for a strong point source in R-band image. The I-band image is more compact. The sources show variability in the near IR, with a very red observed colour during IR maximum.

A strong HI 21 cm absorption is detected with a width of 100 km s$^{-1}$. Its contours extend to the east and consistent with emission from a large-scale jet. Milli-arcsec scale (mas) structure contributes about 70% to the total flux density at 851 MHz and about 85% to the peak surface brightness. It is likely that HI covers most of the milliarcsec jet structure.

*Compact steep-spectrum sources (CSS)*
Some of the CSS sources are young. Their small sizes mean that radio observations probe the dense nuclear environment. Radio depolarization is also common. Evidence for jet-ISM interaction in CSS's is strong, including bent, knotty jets and aligned optical emission (de Vries et al. 1997).

CSS's differ optically from the larger sources. There are very red optical continua, large Balmer decrement, and large narrow-line equivalent widths, which are suggestive of reddening.

Associated absorption by enriched HI is exceptionally common in CSS's. These quasars share many similarities with high-redshift radio galaxies.

*Mg II absorption*
The population of QSO's selected by the presence of MgII is indistinguishable from the population selected by the presence of a Lyman limit break. Mg II is known to trace H I gas, that is optically thick at the Lyman limit. They have velocity spread of ~ 70 km/s. Surveys of Mg II absorbers have detected galaxies, which are seldom fainter than 0.1 $L^*_K$ ( at the redshift seen in the absorption).

In fact, a relatively luminous galaxy, within ~40kpc of the QSO line-of-sight, appears to be a prerequisite for the detection of MgII absorption. The spatial distribution of the absorbing gas surrounding intermediate redshift galaxies is not smoothly varying, and the velocities of the gas clouds can not be described by a single systematic kinematic model.

The absorbers with W(MgII)> 0.3Å have generally been interpreted as material infalling into the halos of the normal >0.1L* galaxies. Absorption profiles arising in 0.4< z<1.0 galaxies , exhibit a rich variety of sub-component structure and kinematic complexity.

*Characteristics of the absorbing galaxy*
Typical MgII absorbing galaxy, at intermediate redshifts, is characterized by an inner region of radius ~ 15 kpc, that gives rise to damped Ly-alpha lines, together with a region extending to ~ 40 kpc, that produces MgII absorption lines and Lyman limit break, and an outer region extending to ~ 70 kpc that is less shielded from extragalactic background radiation and produces absorption lines in higher ions, such as CIV.

*Low ionization gas and morphologies of the absorbing galaxies*
The absorptions are believed to arise in the low ionization gas associated with a population of normal field galaxies, that exhibit little to no evolution in the rest frame(B-K) colours since redshift ~ 1.0. Galaxies, selected by the presence of MgII absorption, have wide range of colours - from late-type spirals to the reddest ellipticals, and their $L_B$ and $L_K$ luminosity functions are consistent with the local luminosity functions, in which morphological types later than Sd are excluded.

*LSB/ dwarf galaxies*
At lower redshifts, Ly-alpha clouds are more directly associated with low surface brightness and/or dwarf galaxies, or with the remnant material left over from the formation of galaxies and/or small galaxy groups.

If the weak Mg II systems are closely associated with LSB and/or dwarf galaxies, there, in fact, may exist a narrow redshift range at z ~ 1, over which a break from UV background to local type stellar dominated photoionization occurs, as the UV background falls below a critical level. Since the iron-group enrichment is gradual, such a break could be inferred from a rapid jump in the fraction of weak Mg II systems having strong Fe II absorption.

Relative to the redshift z~ 2.5, the metagalactic UV background flux at z<1 is reduced by a factor of 5. This indicates that the IGM ionization conditions may have evolved so that low ionization species, specially the resonant Mg II doublet and several of the strong FeII transitions, are detectable in Ly-alpha clouds.

*FeII and MgII abundance*
During the star formation epoch Type II SNe disperse alpha group elements such as silicon and magnesium, over short time. Following the star formation peak, the chemical build up of iron-group elements ( a SN Type Ia process) gradually increases over longer time, and the Fe/Mg abundance ratio rises.

Most of the local starbursts have FeII/HII in the range 2-6 (The FeII is a measure of the "supernova" rate.) The nebular H-emission can be used to predict Type II supernova rates, because it gives a measure of the number of massive stars which are presently ionizing the gas. The total number of Type II supernovae is at most a factor 2 lower than the total number of Type Ia supernovae, but is spread over a timescale about 30 times shorter. In a star-forming region, the SN Type II rate is about one order of magnitude larger than the SN Type Ia rate, as observed for late type spirals. In this case the FeII emission is produced mainly by shocks from SN Type II and is a measure of the star formation activity in the recent past.

Observations show that in the galactic ISM, iron is depleted by almost an order of magnitude more than magnesium. But the relative iron depletion is not so severe in the halo. Therefore, for a given abundance pattern, the gas-phase iron to magnesium ratio is higher in the halo than in the disk. If the gas in the halos is continually recycled then the iron to magnesium abundance ratio should increase with decreasing redshift, as the Type Ia iron enrichment is cycled into the halo from the disk. The two effects- dust depletion, and stellar evolution- work in opposite directions in their effects on the iron to magnesium phase abundance ratio. Generally the net result is that iron is enhanced in the disk material.

If the alpha group elements are enhanced relative to Fe group, then the chemical enrichment is dominated by Type II SNe. If a given Ly-alpha cloud is measured to have Fe/H >-1 and alpha/Fe group abundance ratios approach solar properties, then one might infer that type Ia SNe have played a role in the chemical enrichment.

A type of extended gas rich object, that is seen to have Fe/H>-1 and even >0, is the class of the giant low surface brightness galaxies. At low redshift the general population of low surface brightness galaxies outnumber high surface brightness galaxies by a factor of at least two.

Recent discovery of saturated MgII doublet, associated with the z=0.072 dwarf galaxy, suggests that the star formation in these objects may directly govern their gas cross section. If so, active star formation dwarf galaxies could contribute to the overall metal line absorption cross section. One would then expect that the abundance pattern arising from a bursting dwarf will be alpha-group enhanced. It is likely that UV ionizing flux from the newly formed O- and B-stars would contribute to the ionization conditions in the absorbers.

*Interactions*
Hubble space telescope images of z~1 reveal diversity of morphological types including small compact galaxies , and apparently normal late type spirals, and give evidence of a high frequency of merging events in the past, and for interactions of small, gas rich galaxies with each other, or with larger companion galaxies.

Observations of nearby galaxies show that the characteristics of Mg II absorption lines depend on a wide variety of mechanisms. The large velocity range, which the absorption covers, is reminiscent of absorption lines from the High Velocity Clouds in our galaxy (Bowen et al. 1996). One of the explanations of the Milky Way HVCs is that they are the result of interactions between the Galaxy and neighbouring satellites. Moreover, many MgII clouds have ionization structures in which C IV surrounds their lower ionization Ly-alpha -MgII cores ( Churchill 1998).

One can not rule out the idea that some fraction of the observed systems are dwarfs, that undergo episodic self-regulating star formation, resulting in the blow out of gas to larger radii over a Hubble time. It is difficult to reconcile a large contribution from this latter picture due to an observed correlation of the absorption cross section with the identified galaxy K-luminosity (Churchill et al. 1996), which suggests that the galactic mass has some connection with a galaxy's ability to organize the tidally stripped, accreting or infalling material. The galaxies, identified with MgII absorption, are indeed associated in some way with the process, that gives rise to the gas.

*PKS 0454+039*
HIRES/Keck spectrum of PKS 0454+039 for MgII absorption in Ly-alpha clouds show two clouds at z=0.6248 and z=0.9315. The cloud at z=.93 is inferred to have super-solar metallicity 0.5<Z/Z-solar<1.5. There are a few diffuse extended LSB galaxies over the impact parameter 20- 60 kpc (if they have z=0.93).

The two clouds are unique with respect to Ly-alpha forest at large. Their high Fe/H and iron-group enhancement, imply that their environments have been influenced by Type Ia SNe yields. It could be that the absorbers are associated with galaxies. However, there are no candidate of high surface brightness object in the field. The dwarf galaxies of roughly <0.01L* at zero impact can not be ruled out.

*Clustering( Yamada et al. 1997)*
La Franca et al. (1998) have found an evidence for an increase in clustering with increasing redshift for QSO's with $0.3 < z <2.2$. Furthermore, the most luminous objects in the high redshift universe seem to exhibit unusually strong spatial clustering.

A clustering of very red objects in the vicinity of a quasar at z=1.1 has also been observed. Their optical-NIR colours and magnitudes are consistent with brighter cluster galaxies, formed at higher redshift and observed at z=1.1. There are several emission line galaxies near the quasar, and observations strongly suggest that the red objects are old galaxies. Many galaxies, found near 1335.8+2834, are fairly luminous (>0.5 L*). An excess of ~ L* galaxies has also been found near the radio-loud quasar at z ~ 2.

*GRB's and QSO's*
The difficult problem of making QSO model of the gamma-ray bursts is how to explain the extremely short time scale of variations. In the radio band, one can explain the variations of a few weeks by models that involve the relativistic jets. However, one has already problems to explain the optical and the higher frequency variations in the time scale of minutes. The data from the best observed blazars, OJ 287, PKS 2155-304 and Mkn 421, show that the regions of the flares are the same. The regions become continuously more compact as the frequency grows. Longer wavelength flares are usually recurring events. The amplitudes of the flares become larger when the frequency grows. In the radio, the flux increases by a factor of two. It can be ten or more in the optical region. In the UV and X-rays the trend is the same. Similar estimates for the gamma-rays will produce a factor of more than five hundred. This means that the source brightens about seven magnitudes during a gamma-ray flare, and one may see a normally weak source easily with Compton experiments. If the peak of the flare lasts only a few seconds one may detect a GRB.

**III.2. BL Lacs**

Great similarity between BL Lac spectrum of PKS 0138-097 and the optical spectrum of GRB 970508 has been observed. This similarity and the abundance of BL Lacs amongst known bright gamma-ray sources point to correlation.

The BL Lacs are one step down in the luminosity from the quasars. They are even rarer objects with the following attributes:
1) near absence of emission lines and absorption lines in the central source
2) strong and rapid variability in radio, IR and optical bands
3) nonthermal continuum with a peak output in the IR, but extending at least from the radio through the optical
4) strong and variable polarization in at least one band

Many BL Lacs can be described as bright point sources in underlying galaxies. Only a few associations of BL Lacs with groups of galaxies are known. The space density of BL Lacs, with extended radio emission > $10^{29}$ erg s$^{-1}$ Hz$^{-1}$ sr$^{-1}$, is > $1.7 \times 10^{-7}$ Mpc$^{-3}$, while the space density of similarly powerful radio galaxies is $10^{-5}$ Mpc$^{-3}$, indicating that 2% of such galaxies have the orientations that can be observed as BL Lacs. The radio galaxy cores, if moving with Doppler factor > 4, can appear to be such intense BL Lac nuclei.

*Radio*
They possess low luminosity extended radio structures that often resemble FR 1's. Their overall spectral energy distributions and luminosity functions are consistent with the Doppler boosted FR 1 energy distribution and luminosity functions. The multi wavelength properties of the X-ray selected BL Lacs are well described by the beamed FR 1 model. However, many of the radio selected BL Lacs have an overall energy distribution similar to the flat radio-spectrum quasars with weak and broad emission lines, and extended radio structures that are more luminous than FR 1's.

Radio spectra of BL Lacs are flatter than QSO's and they remain optically thick upto about 90 GHz (QSO's are optically thin above 10 GHz). About two-thirds of the flux come from within a region of 0.3-0.5 mas, and an elongated jet-like structure of 1.5 X 0.5 mas. BL Lac OJ 287 shows a circular core that emits >80% of the energy from a region of 0.3 mas diameter, and a halo of at least 4 mas around it. This core-halo morphology occasionally shows jets. The estimated linear sizes of the halos, or the jets are 20-300 kpc. The total radio power lies in the range $10^{41}$-$10^{44}$ erg/s. Its spectral indices are consistent if the lobes are being viewed end on.

However, a fraction of the BL Lacs show more extended structures. For 0219+428 (3C 66A) and 2155-304, the estimated linear scales range from < 0.3 kpc to 500 kpc. BL Lacs, that are also detected in the X-rays, indicate that, in at least two cases (0453+844 and 1803+784), some of the extended radio emission are coincident with the X-ray knots.

*Variability*
Many BL Lacs show variations in radio band over just a few weeks, while QSO's take months or longer. Apart from that the variations in polarization are far more significant in BL Lacs. At 3.7 cm, the magnitude of polarization ranges upto 16%, which is not as large as in the optical, but significantly greater than in QSO's.

*Optical, infrared and ultraviolet*
They have optical emission one to two magnitudes brighter than the equally strong QSR's. Polarization in IR and visible are correlated, as are the timescales of variability. The extremely powerful and the rapid fluctuations, that are observed in 0235+164 and 1308+326, are inconsistent with an ordinary synchrotron emission. The extremely rapid variations in OJ 287 at 1.25 mm demand relativistic, or at least highly anisotropic emission. Changes of upto 1 magnitude, with rise and fall timescales of less than one minute, are present in BL Lacs.

A paucity of ionized gas could explain the rarity of emission lines.

*X-ray observations and correlation*
They are moderate X-ray emitters. The spectral indices cluster around -1, and the hard X-ray spectra show a lack of low-energy cutoffs, implying nonthermal origin. Large scale X-ray knots, associated with radio lobes, and the alignments with respect to the significant optical polarization seem to exist . They have X-ray spectra steeper than QSO's and Seyferts.

Proposals have been made that as the angle between the observer's line-of-sight and the jet axis decreases, sources may switch from a radio-quiet QSO to OVV, or BL Lac, while the radio galaxies, pointing towards us, would appear as BL Lacs. The number density of BL Lacs with $L_X$ (0.5-4.5 keV) > $10^{44}$ erg $s^{-1}$, is approximately equal to that of the optically selected QSO's, or Seyferts with comparable X-ray luminosity.

*MgII absorber (Stocke & Rector 1998)*
Total number of MgII systems in the 1Jy radio-selected BL Lacs sample stands today around 10. Some of the absorbers might be intrinsic to BL Lacs, or there may be a correlation of absorbing gas in the foreground, and the nearly featureless spectra. Four of the ten systems occur in objects that lack emission line redshifts.

BL Lac 0454+844 shows Mg II doublet at z=1.34. It is the most distant known in the 1 Jy sample. The speed of its jet is superluminal.

BL Lac 2029+121 has emission lines of C IV, C III and Mg II at z=1.215 with foreground Mg II, Mg I , Fe II/Mn 2600, 2606 Å and Fe II 2382Å absorption at z =1.117.

BL Lac 0138-097 too possesses Mg II absorption doublet. Emission lines of weak Mg II and O II correspond to a redshift of z=0.733.

There are ample evidences that the beamed FR 1, for low-z BL Lacs, are objects which do not contribute to the Mg II pathlength. The high-z BL Lacs have higher radio power levels than FR 1's, and weak quasar-like emission lines in the optical spectra.

**III. 3. Blazar**

One of the *long-term variable BL Lac* (variations by a factor of two or three on a time scale of several months) is a superluminal quasar PKS 0420-014. Its strongest outburst at the optical wavelengths was at a time coincident with the highest gamma-ray emission. During the optical flare( which shows delayed radio counterpart) the gamma flux was larger than any other epoch.

Many sources with large gamma-ray fluxes also show *intra-day variability*. Outbursts occur at lower energies too. Major radio outbursts and gamma-ray emission are correlated in such sources. Usually the epoch of the gamma-ray detection is followed by (with a delay of several months) an outburst in the radio frequencies. The large fraction of the IDV's are above 100 MeV.

The *rapidly variable objects* are usually optically bright. The flares in the gamma-rays and the optical wavelengths have similar shapes, but are separated in time. The optical flare precedes the gamma-ray maximum by several hours. Variations on time scale of a few hours imply a diameter of a few $10^{15}$ cm, and hence, brightness temperatures, at 5 GHz of upto $10^{18}$K. Simultaneous optical and radio observations show correlation. The radiation may originate in the knots, that are observed in the radio range to be moving outward at superluminal speeds. The spectrum can be represented by a power law with a differential spectral index of 1.6 and 2.6 (the average index is 2.1 and is independent of z within uncertainties).

*Violently variable ones* show quick and significant luminosity variations, particularly, in the optical and the infrared domains. Sometimes they are referred to as "**Blazars**". While they exhibit emission lines, their extreme variability and relatively high polarization match BL Lacs. The most quickly changing objects are 3C 345 and 3C 279, who exhibit apparent superluminal motions.

Only these blazars, that are radio-loud, and have highly polarized optical emission and flat radio spectra, show high-energy gamma-ray emission. The optical polarization is large and variable in both magnitudes and angles. The significant number of these blazars are found at high redshifts.

Their gamma-ray luminosity generally exceed the luminoisities at other wavelengths by about two order of magnitudes. Their spectra may be represented by a single power-law. Though most sources with large gamma-ray fluxes are intra-day variable. The optically bright ones show more rapid variations. The optical flares, that precede the gamma-ray maximum by several hours, show delayed radio flares by several months. When faint, the optical spectra of blazar show emission lines like normal QSO's, but during outburst the emission lines are washed away as in BL Lacs.

The flares in the gamma and the optical wavelengths have similar shape. Optical and radio observations also show correlation. The emissions seem to originate from knots in jets.

*AO 0235+164*
Arecibo Occultation source, AO 0235+164 is a highly variable radio, x-ray and gamma-ray emitting blazar (Burbridge et al. 1996). It has optical variability with an amplitude of five magnitudes, and it looks like a stellar object with a nebulous extension to the south of the nucleus.

*Radio observation of AO 0235+164*
The radio variations are correlated over at least a factor of 50 frequency range ( from 0.3 GHz to 15 GHz), as expected from an expanding outburst propagating towards lower frequencies with diminishing amplitudes (O'Dell et al.1988). The low frequency variations are an extension of the high frequency variations, and seem to be intrinsic to the source itself. The shortest rise and fall time scales are of weeks duration for GHz frequencies and months for lower frequencies. The typical duration of an enhanced activity is about a year. The source is transparent at high frequencies and possesses a flat spectral index ~ 0.3. The total flux density stays between 1 Jy and 3 Jy over a frequency range of at least 300. Above 10 GHz and upto 30 GHz, the spectrum is noninverted and the variations seem to be simultaneous. Comparison with millimeter observations suggests that this behaviour continues at least to 300 GHz. AO 0235+164 is very compact, being nearly unresolved on milliarcsecond scale from 0.4 GHz to 22 GHz. The highest resolution 22 GHz map finds that 1.3 Jy originates within a core of maximum diameter $0.1 \pm 0.03$ mas.

VLBI experiments at 6 cm indicates (Jones et al. 1984) that the jet in AO 0235+164 is undergoing a very rapid precession.

*Millimeter and IR observation of AO 0235+164*
In millimeter and infrared (Rieke et al. 1976, Elias et al 1978 and Landau et al. 1983) the spectrum is flat over a range of at least $10^5$ in frequency. However, in the near infra-red and optical regions the spectrum is quite steep.

*Optical*
Its V-mag varies usually from 17.5 to 19.5. The largest optical variations have been as much as 2.25 mag in three days (Webb & Smith 1989). The variations seem to be correlated at optical and radio wavelengths.

It consists of at least three spectral line systems: One weak emission line system of Mg II l2800, [Ne V] l3426 and [O II] l3727 at redshift of 0.94, and two absorption line systems at redshifts of 0.54 and 0.85. The z=0.94 system shows hydrogen lines $H_\delta$ and $H_\gamma$. The equivalent width of the $H_\gamma$ line, which is 7.2Å in the rest frame of AO 0235+164, exceeds the 5 Å limit for an object to be included in the 1 Jy BL Lac sample of Stickel et al. (1991). The z=0.94 system shows broad permitted Mg II and narrow forbidden lines. The spectrum is comparable to a small number of other BL Lacs with nuclear line emission. The line intensities vary over a time scale of a few years (Cohen et al. 1987). The faint nebulous extension, which is a "companion" galaxy, has narrow emission lines of [O II], [O III] and $H\beta$ at redshift 0.542 with R magnitude of 20.9. But the host galaxy of the system at redshift 0.94 has remained undetected. The "companion" appears to possess two curved plumes, or arms on the opposite side of a core. It is an AGN surrounded by faint nebulosity, and shows spectra that has broad absorption lines displaced to the short-wavelength side of C IV, Si IV and N V emission lines, like some of the less extreme BALQSO's. The emission lines of Ly$\alpha$, C IV l1549, and C III l1909 have broad emission wings, characteristic of a QSO. The Ly$\alpha$ emission has a sharp central spike, but no broad absorption. The C IV is asymmetric: The red (low velocity) side is steeper than the blue, and the blue edge of the trough shows possible structure. The C IV absorption extends from roughly 1500 to 6000 km/s. The spectrum of the "companion" is similar to the spectra of the majority of BALQSO, and the strong absorption features are all high ionization metal lines. It is a very active object ejecting $N^{4+}$, $C^{3+}$, $Si^{3+}$ at high velocities into the intergalactic medium. The Seyfert 1's and the narrow emission line galaxies with [O III]/Hb $\leq$ 3 and emission line FWHM$\leq$ 300 km/s, have [O II] luminosity 5-10 times smaller than the total [O II] luminosity observed at the "companion" and the intervening galaxies at z=0.525. The closest match of the "companion" galaxy is Markarian 833.

There are several extremely faint ( R~ 25) pairs of objects near AO 0235+164 which could be part of z=0.851 or z=0.94 group (Nilsson et al. 1996).

*X-ray observation of AO 0235+164*
It was observed by ROSAT between 1993 July 21 and 1993 August 5, roughly, every 3 days. The flux varied significantly from one exposure to another, with the highest count rate being 4.7 times than that of the lowest. The shape of the light curve showed a rapid rise and a more gradual decay. The lowest flux point corresponded to hardest spectrum.

The source was softer when brighter and highly variable, and harder when fainter and relatively constant. This is in disagreement with a synchrotron model, where the spectrum steepens as a result of the decreasing lifetime of the radiating particles with increasing energy, expected in the case of the synchrotron cooling. One may be driven to conclude that there are two separate spectral components (Madejski et al. 1996), while the drop in the intensity of the soft component uncovers the hard component. As to the origin of the soft component there is a circumstantial evidence of the high-energy tail of an electron distribution, which radiates via synchrotron process.

The soft X-ray absorption observed in ROSAT and ASCA data is primarily attributable to helium, carbon and oxygen.

Given the superluminal expansion, if the source of the X-ray emission, appears from behind a partially absorbing cloud, and moves a tenth of a mas, it would correspond to a sampling path shifted by ~ 1 pc at z=0.542, which is comparable to the size of a molecular cloud structure.

*Gamma-ray observation of AO 0235+164*
ERGET data show strong GeV gamma-ray emission (Hunter et al. 1993) with a flux variable from observation to observation (von Montigny et al. 1995) but no rapid variability on the timescale less than a week was observed.

The other ERGET observations were made for two weeks starting on 1994 February 17. At E >100 MeV the source was detected at the level of 14 x $10^{-8}$ photons $cm^{-2}$ $s^{-1}$, which was significantly lower than what was observed by Hunter et al. The data was consistent with a power-law spectrum with $\alpha$ = 1.0.

*Flares of QSO 1156+295(4C 29.45)*
In its most dramatic outburst, the optical output of 4C 29.45 went up by 5 magnitudes during a few months in 1981. During that peak it had a luminosity more than 1.5 x $10^{48}$ erg $s^{-1}$, which places it among the 10 brightest sources known. Optical flickering of 0.1 mag over 20 min timescales has been seen. The optical polarization was large (upto 29%), which was variable in both magnitudes and angles. There was a large radio flare in 1979 and a huge release in energy in radio band that began a few months after 1981-optical flare. This was seen first at 90 GHz, and later at lower frequencies down to 1.4 GHz. Radio variability over a few days was found near the peak of the outburst. When it was faint in optical spectrum, it appeared like a normal QSO, but during the outburst its spectrum became nearly featureless like a BL Lac. The slopes in IR, optical and UV do not change much despite the huge flux variations. Models based on synchrotron self-Comptonization require the source to possess a bulk relativistic motion with Doppler factor 2-8, a linear size of 0.01 pc, and a magnetic field between 0.1 and 30 Gauss.

# IV. Jets and gamma-ray bursts

**IV.1. Gamma-rays from AGN**

The AGN's that emit gamma-rays, are associated, in particular, with compact core flat spectrum radio sources showing evidence of superluminal motion. There is also a correlation between the radio brightness and the gamma-ray brightness, because the brighter flat spectrum quasars tend to be the AGN's at gamma-ray energies.

In fact , all the EGRET AGN sources are radio-loud flat spectrum sources, which suggest that the gamma-rays are produced by jets, that are beamed in our direction, in a similar fashion to the radio emission. Additional support for the beaming of the gamma-rays from AGN's comes from the fact that several of the gamma-ray loud AGN's have large redshifts and they are emitting at an extraordinary rate ($> 10^{49}$ b ergs s$^{-1}$, where «b» is the beaming factor. In view of the continuing rapid variations observed on timescales < 1 h, the beaming of gamma-rays is required in order to avoid a large induced optical thickness, due to the pair production, that are emitted from a region whose dimensions are restricted by causality to satisfy    $R < c\Delta t/(1+z)$.

As a consequence of the relativistic bulk motion, the observed burst duration for a source at redshift z is reduced by a factor $\delta(a+z)$, where $\delta = [\gamma(1-\beta \cos\theta)]^{-1}$ is the Doppler boost factor, $\gamma$ is the Lorentz factor, $\beta = \sqrt{(1-1/\gamma^2)}$, and $\theta$ is the angle between the relativistic jet and the observer's line-of-sight. The observed luminosity Lo is related to the luminosity L in the comoving frame by Lo= $[\delta(1+z)]^3 L$. In the absence of relativistic beaming  the source size is limited by the burst duration ( $R < c\Delta t$), but it can not be so compact that all the gamma-rays are absorbed in pair producing photon-photon interactions. For a typical gamma-ray burst, values of $\delta > 200$ are required to avoid significant attenuation of high-energy gamma-ray photons in the source region. The characteristic VLBI jets are typically in the range $\delta \approx 3\text{-}20$ and are observed at distances greater than 1 parsec from the central engine.  It is not unreasonable to postulate that the bursts of the particles may be accelerated to values of  $\delta > 200$ during episodic accretion events onto a supermassive black-hole, as well as in the process leading to the formation of massive black-holes. The particle burst is composed of either electron-positron pairs, or protons that are accelerated away from the black-holes.

*Synchrotron self-Compton process*
In the case of the electron-positron pair plasma the gamma-rays can be produced by the synchrotron self-Compton process. However the Compton drag on the plasma, due to the thermal and nonthermal photons from the accretion disk and the material, that surrounds the central engine, yields a very short radiative lifetime with most of the energy being depleted within 20 $R_S$ ( where $R_S$ is the Schwarzschild radius). If the column density in the line-of-sight in the cocoon of gas is greater than $10^{24}$ cm$^{-2}$, the source will be optically thick to photons of energy up to 30 keV. The absorption of X-rays and the rapidly decreasing value of $\delta$ can create the hard to soft spectral evolution. The delayed X-rays may originate after the relativistic plasma has traversed most of the gas but these X-rays will be beamed into a larger angle because $\delta$ has been reduced to values of ~ 50. The different solid angle effects will yield a population of X-ray bursts without associated gamma-ray emission.

In the first light-hours of the life of the plasma, the compactness is high enough to make it opaque to gamma-photons owing to gamma-gamma pair production while pair annihilation is  negligible. Most of the gamma-photons, produced by the annihilation event, will be eventually reabsorbed, and go in creating new pairs. Between 1 light-day and 1 light-week, the accelerated beam will become transparent to gamma-photons though the radius of the beam will still be small enough to allow pair annihilations to take place. When most of the low-energy pairs have been annihilated, the gamma-ray emission will take place. After a few light-weeks  the beam will become transparent in the millimeter range, producing VLBI components by synchrotron emission from the highly relativistic pairs.

*Inverse Compton scattering*
Inverse Compton scattering of soft photons from the highly relativistic electrons in the jet can also produce gamma-rays, which are beamed along the jet axis. These soft photons can be self-generated by the electrons in the jet, or emitted by nearby stars in a dense stellar region, through which the jet propagates. If the Lorentz factor is less than $m_e c^2/E_o$, where $E_o$ is the energy of the soft photons, the radiation field of the soft photons is transparent to the gamma-rays produced by the Inverse Compton scattering. This is also the condition for a small optical thickness for self-absorption of the emitted gamma-rays. For Inverse Compton scattering, the limited radiation is beamed into a cone with an opening angle $1/\gamma$ ( where $\gamma$ is the Lorentz factor) along the jet axis. This keeps the c.m. energy for the pair production below $2m_e c^2$- the threshold c.m. energy. Moreover, the beaming and energy boosting of the optical photons in the direction of the observer by Compton scattering from highly relativistic electrons in the jets, may explain why the gamma-ray emission along the jet is more enhanced than the optical emission in the direction of the observer.

**IV.2. Gamma-ray production from jets**

*Model for jet formation and collimation*
Magnetic fields are necessary for the formation of relativistic jets. Dynamo action in the inner accretion disk can create a rotating magnetosphere, which is filled with plasma. Along open flux surfaces the plasma can be accelerated to relativistic speeds and collimated outside the light cylinder near the last stable orbit around a rapidly rotating Kerr black-hole. The plasma moves in the magnetized jets in such a way that the total energy and the total angular momentum along a magnetic flux surface are conserved.

Initially the energy and the momentum in the jet are carried away by the electromagnetic fields. However, as the jet propagates from the centre, there is a steady conversion of electromagnetic energy into energy carried by the electrons and the positrons. The transition from electromagnetic to particle dominance occurs in the vicinity of the annihilation radius at the end of the inner jet. At small radii, within $R_{ann}$, the bulk flow is steadily accelerated, so that the Lorentz factor steadily increases. When this factor exceeds a critical value, background X-rays can be scattered as gamma-rays with energies above the threshold energy to pair production on the ambient X-rays, thereby giving rise to copious pair production, which will increase the inertia of the jet, as well as create a greater radiative drag on the outflow, and eventually limit the jet Lorentz factor at a modest critical value. The details of this limiting process depend on the prescription for the particle acceleration and the spectrum of the external radiation field. If the superluminal motion reflects the speed of the bulk flow, rather than that of a shock propagating down the jet, the correlation between the gamma-ray breaks and the superluminal speeds is expected.

When a compact object turns around a massive black-hole its gravitational interaction perturbs the accretion disk and induce instabilities. The perturbation will produce a precession of the non-relativistic jet via the magnetic field lines which are frozen in the plasma of the disk. The Alfven waves, generated in the jet due to the perturbation, will extract $e\pm$ component which propagates relativistically in the processing beam. The $e\pm$ component radiates anisotropically in the X-ray and gamma-rays. With the bulk Lorentz factor about 10, the maximum value of the Doppler beaming factor is about 20. Due to the rotation of the $e\pm$ component the beaming factor varies strongly and produces quasi-periodic variability. Supposing that the redshifts are ~1, the bulk Lorentz factor is about 10 close to the nucleus, the magnetic field in the beam is about 100 G, and the densities of the $e\pm$ are about $10^9$ $e^-$ $cm^{-3}$, the relativistic Alfven speed of the beam is 0.71c. With this propagation speed of the Alfven wave and the speed of the relativistic component about c, the component will escape the wave front after a characteristic distance «b» from the nucleus, and will move straight on in the unperturbed beam. So the Doppler factor will show strong variations at the beginning and become constant when the component moves straight on.

Knots: Magnetic effects of the disk may inject plasma bubbles into the overall axisymmetric flow. These bubbles are then dragged along with the underlying continuous flow so that the specific angular momentum of the bubble stays constant along the flux tube. Since the outflowing jet plasma carries away the angular momentum, a bubble in the jet will also rotate. Under the expansion of the jet the local rotation period will increase.

Flaring: The spectral flux, observed from an unresolved knot, moving in a collimated jet with constant opening angle, is strongly amplified by beaming. Since the Doppler factor is a function of time, this flux is in general time-dependent. But due to the rotation of the bubble, as the knot moves along a magnetic surface with a constant opening, the beaming factor can strongly vary and lead to a quasi-periodic flaring. Short periods of a few days are observable due to this effect.

The injection of such bubbles into rotating relativistic jets may explain the intra-day variability in the optical and the radio emission of the compact sources. The rapid variability is a consequence of the time-dependence of the angle between the line-of-sight and the velocity vector of the knot. The effect is mostly visible for non-axisymmetric structures in the innermost part of the jets. The shape of the light curve is strongly dependent on the opening angle. If the varying opening angle is taken into account, almost every kind of light curve can be fitted with this model. However, the trajectory of the bubble has to be very close to the axis in order to get the short timescales and powerful beaming. The time evolution of the relativistic particles will also influence the overall shape of the lightcurve.

*Lighthouse effect of relativistic jets (Camenzind & Krockenberger 1992)*
There is a correlation between the flux outbursts in IR-optical region and the birth of a new superluminal knot in the parsec-scale jet.

During the optical activity of 3C273 in 1988, preceding the ejection of the superluminal knot C9, quasi periodic fluctuation was observed with a fundamental period of 13 days. The actual birth of the knot coincided with a strong X-ray flare. It is believed that the light cylinder of the innermost part of the disk magnetosphere is extremely compact, the knots are injected into the jets at a radius ten time the radius of the light cylinder with a typical magnetic field strength of 1 Gauss, while the knots rotate with the jet plasma at a fundamental period of 660 days. Due to the beaming effect, this period appears shortened to 13 days. If the opening would be much larger the flares would be damped and no quasi periodicity would be visible. After a few revolution the spectrum of the knot moves into mm-region. At a distance of a few milliarcsec from the core it will be washed away by the diffusion of the relativistic particles.

Similar optical flares observed for the outburst of 3C 273 in 1988 with time scales of a fraction of a day are found in some BL Lacs. A few of the sources show quasi-periodicity in the optical as well as in the radio regime. The variability are more pronounced at higher frequencies with rapid timescales. The timescales for changes in brightness are usually longer at radio frequencies.

*Anistropic inverse Compton scattering model in relativistically beamed emission blobs ( Dermer et al. 1992)*
As the blobs erupt from the nucleus, relativistic electrons with quasi-isotropic distribution, in the rest frame of the blob, produce incoherent synchrotron radiation, which, because of the blob's motion, is focused into a cone about the jet axis in the observer's frame.

The accretion disk and the core emit photons in the UV and X-ray range, that illuminate the blob from behind (the blob is optically thin to Thomson scattering along the jet axis, which is necessary to produce the highly polarized radio emission). Superluminal sources should scatter most energetic photons at the observer's direction, at which the apparent SL velocity is greatest. Thus the SL radio sources should be the strong gamma-ray emitters. But if observations are made directly down the symmetry axis of the jet, one should see very little X-ray and gamma-ray emission.

The importance of the anisotropic inverse Compton scattering depends on whether the synchrotron and the synchrotron self-Compton losses dominate the total energy loss rate of the electrons in the jet. Moreover, the energy density of the central source photons should be greater than the magnetic field energy density at the emission site.

This process would be important if there are knots of high magnetic field in the jet, near which the electrons are accelerated to emit radio waves, surrounded by regions of low magnetic field, into which relativistic electrons diffuse and Compton-scatter the soft photons.

By upscattering UV and X-ray photons far from the nucleus, it avoids the gamma-gamma pair attenuation of >100 MeV photons from a highly variable source (such as 3C 273). Correlated multiwavelength variability could be associated with the outflowing blob.

The model also predicts that the gamma-ray intensity should be positively correlated with the velocity of the superluminal blobs, and the proximity of the blobs to the compact cores.

## IV.3. Gamma-ray bursts from jets

Except for the timescales, striking similarities exist between the high-energy gamma-ray emission from such AGN sources and the gamma-ray bursts: Both show rapid variations in luminosity, a diversity of light curves, a complex time structure with exponential temporal autocorrelation functions, and a power-law spectrum $dN\gamma/dE \sim E^{-\alpha}$ with approximately the same powers $1.5 \leq \alpha \leq 2.5$, that extends to high energies. Their apparent luminosity in high-energy gamma-rays are typically larger by more than two orders of magnitude than their observed luminosity in other bands of the electromagnetic spectrum. For GRB's, like AGN's, the peak luminosity, within the solid angle in which the emission occur, are highly super-Eddington for compact objects, which can produce the short time variability. Both gamma-ray spectra do not show a cut-off due to self-absorption via the pair production process $\gamma + \gamma \rightarrow e^+ e^-$.

The gamma-ray bursts from jets have been discussed by several authors ( Roland et al. 1994, Dermer & Schlickeiser 1994, Krolik & Pier 1994, Brainerd 1992, Mc Breen et al. 1993). However, Shaviv and Dar (1995) claims that Inverse Compton scattering of soft photons, can explain many of the observed features of the gamma-ray bursts, including the duration histogram and the correlation between their temporal characteristics. They assume that the jet to be highly relativistic, the initial photons are < 511 keV, the photon source size is small compared with the interaction length, and the total fluence, needed for detection of the GRB, is independent of the duration of the GRB.

The Inverse Compton scattering from knots will give rise to multipeak gamma-ray bursts, that will look very complex, but possess rather simple power spectra, which seem to be a general feature of the GRB's.

The light curves of the gamma-ray bursters will be sensitive to the bulk Lorentz factor, the precessing angle, the observational angle and the detection limit. The greater the precession angle the sharper will be the gamma-ray peak. The number of the peaks depends on the parameter «b» which corresponds to the distance after which the component escapes the Alfven perturbation and moves straight on. The greater this parameter greater will be the number of peaks observed. A relatively large precession angle will produce an increase of the time scale between these peaks. When the precession angle is $\approx 0.1°$ the light curve will be mostly periodic. The bursters with few peaks and smooth profile correspond to smaller parameter «b» i.e. the ejected component will make only one, or two, rotations before it moves straight on. The peaks will be larger if the precession angle and the Lorentz factor are smaller. When the parameter «b» becomes as small as $10^{-7}$ pc, the light curve will present a single peak with smooth profile. One should observe a sharper decay of the light curve at higher energies.

The high frequency perturbation of the beam will correspond to the frequency of the compact object turning around the massive black hole at almost 3 Schwarzschild radii.

As the typical time scale between the observed peaks is smaller than 10 s the mass of the central object will be smaller than $10^6 M_\odot$. For some gamma-ray bursters the timescale between peaks can be as small as 0.1 s, indicating that the mass of the central black-hole can be as small as $3 \times 10^4 M_\odot$.

# V. Supernovae

## V.1. GRB's and supernovae

A GRB was observed two sigma from the position of the supernova SN 1992ar, discovered in July 1992. SN 1997ef, discovered on November 25, 1997, was within 3 sigma error box with GRB 971115 and GRB 971120, and GRB 970514 was less than a degree away from SN 1997cy, the most luminous supernova discovered. At maximum it had $M_R \approx -21$ with a bizzare spectrum that included broad SN Type Ic-like lines (as observed in SN 1997ef and SN 1992ar), together with a H-alpha line with broad and narrow components.

Recently Galama et al. (1998) reported the discovery of a supernova SN1998bw within the error box of GRB 980425. It was offset from the nucleus of a face-on barred spiral galaxy ESO 184-G82 (z=0.0085). A bright radio source was found coincident with the GRB.

If there exists no hydrogen in the spectra, the supernova is called Type I, and if there is hydrogen, it is called Type II ( Filippenko 1990). A subclass, characterized by i) lack of the 6129 Å absorption trough (thought to be blue shifted Si II 6347, 6371 in normal Type I), ii) preference for galaxies of types Sbc or later, iii) proximity to H II regions, iv) rather low luminosity (typically 1.5 mag fainter than normal Type I), v) distinct IR light curves having no secondary maximum around one month past primary maximum, vi) reddish colours, and vii) radio emission within a year past maximum, is called Type Ib. The spectroscopic appearance of this Type Ib , near maximum,  resembles that of older Type Ia ( a month past maximum).

The optical spectra of  SN 1998bw, that resemble those of SN 1997ef, can not be classified under such scheme. At an early phase SN 1997ef showed broad emission/absorption features, with FWHM about 50 000 km/s. These features evolved rapidly to the red at a rate of about 1500 km s$^{-1}$ day$^{-1}$ during the first week.  At a later time the spectrum resembled a SN Type Ic, characterized by little hydrogen, helium, or the strong Si II line of SN Type Ia. The early spectra of SN 1998bw and SN 1997ef did not resemble a canonical SN Type Ib, or  SN Type Ic.

If the supernova SN 1998bw is powered by the decay of radioactivity, like other Type I's, it is necessary to synthesize and eject $> 0.45$ solar mass of $^{56}$Ni in its explosion. Large $^{56}$Ni mass requires a very massive star and an explosion energy large enough to accelerate this ejected mass to the high velocities and make the light curve peak in only 17 days, which is several times of a typical SN Type Ib. It is believed  that such GRB's will require so massive stars that the neutrino powered mechanism for explosion will fail, leading one to consider stars whose main sequence mass is over 30 solar mass (Woosley et al. 1998).

It is also possible that such explosions of massive stars could be asymmetric and powered by jets. If matter can fall in close to the black hole and come out again, the production of $^{56}$Ni will not be directly tied to the shock energy and pre-explosive density structure of the star. It could be made "convectively" instead.

## V.2. "Failed Type I supernova/ hypernova"

When a core of a massive star collapses, the magnetic field may reach $\sim 10^{15}$ Gauss. Right after this, the collapsed core may have up to $5 \times 10^{54}$ erg stored in its rotation. If a large fraction of that energy is extracted rapidly by a strong magnetic field, the kinetic energy of explosions may reach $\sim 10^{54}$ erg. If the ejected mass is $\sim 10\ M_\odot$, then most of the ejecta will be sub-relativistic.

However, a small fraction may reach relativistic velocities. When the leading relativistic shock will slow down, as it will sweep up the ambient medium, the slower moving matter will catch up from behind and provide a long-lasting afterglow.

## V.3. Wolf-Rayet stars in very young starburst galaxies

All massive single stars with initial masses above a certain limit are supposed to evolve into Wolf-Rayet stars (WR stars), that are the offspring of the most massive O-stars, and whose lifetime is at most $10^6$ years, have burning helium in their centres. They have lost all, or most of their hydrogen rich envelope. The very intense stellar wind is responsible for the dominance of the emission lines in their spectra ( Lange 1990). Most of the WR stars, after exhaustion of core-helium, undergo central carbon, neon, oxygen and silicon burning, leading to the formation of iron-cores, which eventually collapse. The iron-core mass depends on the actual mass of the star at the end of its evolution, which does not change after the core-helium exhaustion. Because of the very high mass loss rate, the final WR mass is much smaller than the corresponding zero age main sequence mass, and in most cases, even much smaller than the helium core mass at the hydrogen exhaustion.

There exist two spectroscopic subclasses of WR stars: One with partial or complete hydrogen burning, and the other with incomplete helium burning. Both subclasses are further divided into subtypes. Late WN stars (WNL) still retain a considerable amount of hydrogen in their atmospheres, in contrast to the early WN stars (WNE), where hydrogen seems to be very rare, or absent. In an evolutionary scenario for the WR stars a WNL phase precedes the WNE phase, because the hydrogen burning convective core shrinks with time. In order to expose ashes of the complete H-burning at the surface (WNE phase), the ashes of the incomplete H-burning have to be exposed earlier (WNL phase). Moreover, because of the He-burning, convective core is smaller than the H-burning convective core, and a WC phase is preceded by a WN phase.

The bulk of the observed WR stars are mostly WNE and WC stars, and most massive WR stars are WNL stars. A 60 solar-mass population I star, which may be typical WR progenitor, is found to end its evolution as a 5.8 solar mass WC star, which is able to perform a supernova explosion after core collapse. The light curves of the exploding low mass WR stars are compatible with observed SN Type Ib light curves (Van den Bergh 1992).

If WR stars are the progenitors of supernovae of SN Type Ib, one would expect the majority of the massive core-collapse supernovae to suffer heavy interstellar absorption. If the progenitors of SN II are less massive, then the locations, at which SN II explode will be weakly correlated with the massive dusty clouds, in which they are formed, and one would expect the mean reddening suffered by SN Type II, to be lower than for SN Type Ib. Consequently the rate of SN Type II discoveries is expected to be less affected by absorption than the SN Type Ib's.

# VI. Starbursts and the epochs of the galaxy formation

## VI.1. Starburst

In the local universe, the high-mass stars are found both in the normal and the starburst galaxies. In the normal galaxies they are distributed in the spiral arms throughout a ~ 30 kpc scale disk, where the supply of gas can sustain the rate of star formation for many Gyrs. In the starburst galaxies the high-mass stars are concentrated in a small region in the galactic centre (~ 100 times smaller than the galaxy as a whole), and they create transient events of star formation with a duration < $10^8$ years.

Within 10 Mpc distance from our Galaxy, 25% of the high-mass star formation is accounted for by less than a handful of starburst galaxies. At intermediate redshifts, the bulk of the excess faint blue galaxies could be due to a dwarf galaxy population, which is experiencing high-efficiency fast evolving starbursts. At higher redshifts the galaxies indicate a high intensity star forming phase.

*Redshift evolution*
Star forming galaxies with spectra broadly similar to those of the present day starbursts have been detected upto z ~ 5. At such early time many of the characteristics of the present day universe were already present. The intergalactic medium was fully ionized, galaxies were enriched in a wide variety of chemical elements, and structures on the scale of the rich clusters were already evident.

*Starburst galaxies in the local universe(Rej & Østgaard 1997)*
In the nearby universe( < 10 Mpc), the four most luminous starburst galaxies are M82, NGC 253, M 83 and NGC 4945 . Together they form 25% of the total high-mass star formation in this region. In fact, the rate of high-mass star formation in the few-hundred-parsec-scale starbursts in M 82 exceeds the rate in the entire disk of the spiral galaxy M 101 , which has a surface area approximately four orders of magnitude larger than the M 82. Thus both in terms of energy production and rate of high-mass star formation, starburst galaxies are highly significant components of the present universe.

In many local starbursts the interstellar absorption lines are significantly blue-shifted with respect to the systemic velocity of the galaxy. The true galactic systemic velocity lies between the velocity of the UV interstellar absorption lines, and the Ly$\alpha$ emission lines. This is due to the outflowing gas, which produces the blue-shifted absorption lines, as well as absorbs the blue side of the Ly$\alpha$ emission line. Thus a purely UV signature of the outflowing gas is a blueshift of the interstellar absorption lines with respect to the Ly$\alpha$ emission line.

*Vacuum UV regime*
The vacuum-UV spectral regime ( 912 -3000 Å) is the energetically most dominant region for the hot stars, that power starbursts. The vacuum-UV spectra of the starbursts are characterized by strong absorption features, which can have different origins like stellar wind, stellar photospheres, and interstellar gas. The resonance lines, due to objects with low ionization potentials (OI, CII, SiII, FeII, AlII, etc.), are primarily interstellar. In contrast, the resonance lines due to high ionization (NV, SiIV, CIV) can contain significant contribution from both stellar winds and interstellar gas. The stellar photospheric lines are usually weak.

The greater is the fraction of the UV, which is absorbed by dust and re-radiated in the far IR, the redder becomes the vacuum-UV continuum. Data strongly suggests that the dust responsible for the vacuum UV extinction, is distributed in the form of an inhomogeneous foreground, or "sheet" surrounding the starbursts. Moreover, the vacuum UV extinction correlates strongly with the bolometric luminosity. Only the starbursts with $L_{bol}$< few x $10^9$ $L_\odot$ have the colours expected for unreddened starbursts, and vacuum UV luminosity that rival the far-IR luminosity.

The starbursts which lie above $L_{bol}$> few x$10^{10}$ $L_\odot$ have red continua and are dominated by the far-IR emission.

Apart from the effect of the dust, the starburst's metallicity is the most important parameter in determining the UV properties: At low metallicity a significant fraction of vacuum UV escapes starbursts, and the vacuum-UV colours are consistent with the intrinsic(unreddened) colours expected for a starburst population. In contrast, at high metallicities (> the solar), 90-99% of the energy emerges in the far-IR, and the vacuum UV colours are very red. This means that vacuum-UV radiation escaping from starbursts suffers an increasing amount of reddening and extinction as the dust-to-gas ratio in the starburst ISM increases with metallicity.

The metallicity dependent strengths of the UV absorption lines, which are stellar photospheric, are generally weak. They include CIII, SiIII, SV, SiIII and FeIII. The strong interstellar lines are optically thick. Their strength is determined to first order by velocity dispersion in the gas. The enormous strengths of the starburst interstellar lines require very large velocity dispersion (of a few hundred km/s). The interstellar lines are often blueshifted by one-to-several hundred km s$^{-1}$ with respect to the systemic velocity, showing that the absorbing gas is flowing outward, feeding the "superwinds".

*Superwinds*
Superwinds are galactic scale outflows. The temperature of the hot outflowing gas in a starburst is considerably cooler (a few to ten million degrees) than it would be expected for pure thermalized supernovae and stellar ejecta (at $10^8$ K). Soft X-ray emission (hot gas) is a generic feature of the halos of the nearest starburst galaxies. The estimated thermal energy of this gas represents a significant fraction of the time-integrated mechanical energy supplied by the starbursts. Superwind transport much of the mechanical energy supplied by the high-mass stars into the IGM (Inter Galactic Medium).

If superwinds carry substantial amounts of metal out of the starbursts, one will see cumulative effects of these flows in the form of a metal-enriched IGM, and/or metal-enriched gaseous halos around the galaxies. Typical MgII absorption line systems at z<1, seen in the spectra of distant QSO's, arise in the metal enriched halos of the intervening galaxies. It is possible that the galactic halos/IGM may be polluted by episodic eruptions associated with powerful starbursts.

*Causes of starbursts*
As massive stars explode, the surrounding gas acquires thermal energy, and the gas temperature rise up to about $10^6$ K. At the same time the gas is polluted with synthesized metals. A shock wave propagates outward as the superwind drives outflow from inside. This outflow collides with the infalling gas, and high-density super-shell is formed. While the gas is continuously swept up by the super-shell, the density increases due to an enhanced cooling rate in the dense shell. Then the intense star formation begins within such super-shell. Subsequent SN explosions accelerate the outward expansion of the shell. Star formation continues in the expanding shell (for about $10^8$ years) until the gas density becomes too low to create new stars. The oscillations of swelling and contraction of the system continue for several times $10^8$ years, and the system becomes settled in a quasi-steady state in $3 \times 10^9$ years. The resulting system forms a loosely bound virialized system due to significant mass loss, and has a large velocity dispersion, and a large core( and stars are formed before most of the gas is fully polluted by metals). Radial distribution of metal abundance in this system has a positive gradient, which is in sharp contrast to the observed negative gradient for massive galaxies. This process of star formation turns out to reproduce the features of dwarf ellipticals.

*Starburst galaxies at very high redshifts( Pettini et al. 1997b)*
The high-z galaxies in the HDF are relatively faint. Their star formation rates are less than 20 solar masses per year. Madau et al.(1996) have estimated a deficit by a factor of 10 in the comoving density of star forming ellipticals at z =2.75, compared to the present day density of L>L* ellipticals. This raises the possibility that giant ellipticals formed at z < 2.5.

Similar to the bright galaxies at the present epochs, the high-redshift galaxies are characterised by strong clustering. HST imaging of z>3 shows that these objects are bright galaxies in an epoch of relatively intense star formation. Most of the z>3 galaxies are characterized by compact morphology, generally having a core ≤ 1.5 arcsec in diameter. The core typically contains about 90-95% of the total luminosity and has a half-light radius in the range 0.2-0.3 arcsec.

However, the star forming galaxies at high-redshift appear to be spatially more extended versions of the local starbursts. The same physical processes, which limit the maximum star formation intensity in nearby starbursts, also seem to be at play in the galaxies at high redshifts. The average surface mass-density of the stars within a half-light radius(~ $10^2$-$10^3$ $M_\odot$ pc$^{-2}$) is quite similar to the values in present day ellipticals.

Powerful starbursts in the present universe emit almost all their light in the far-infrared rather than in UV. This may be also true at high redshifts. It is estimated ( Meurer et al. 1997) that an average vacuum-UV-selected galaxy at high redshift suffers 2 to 3 magnitudes of extinction. If the strong correlation between vacuum-UV colour and metallicity in the local starbursts is applied to high-z galaxies, it would suggest a broad range of metallicity, from substantially sub-solar to solar, or higher values, and a median value of 0.3-0.5 solar values. This is significantly higher than the mean metallicity in the damped Lyα systems. If extinction corrections are applied, the high-z galaxies possess large bolometric luminosity ( ~ $10^{11}$ to $10^{13}$ $L_\odot$). This bolometric surface brightness are very similar to the values seen in the local starbursts. The implied average surface-mass-density of the stars within the half-light radius is also quite similar to the values in the present-day ellipticals.

A typical high-z galaxy with R=24.5 and z=3 has a far-UV luminosity L= 1.3 x$10^{41}$ $h_{70}^{-2}$erg s$^{-1}$ Å$^{-1}$ at 1500 Å ( with Hubble constant h0=70 km/s). At z~3 the UV luminosity exceeds by a factor of 30 the most luminous local Wolf- Rayet Galaxy NGC 1741, which contains $10^4$ O-type stars. Their typical luminosity corresponds to a star formation rate of ~ 8 $h_{70}^{-2}$ $M_\odot$ yr$^{-1}$. This is a lower limit because the dust extinction and a lower age would raise the value. Pettini et al. estimates a star formation intensity ~ 13 $M_\odot$ yr$^{-1}$ kpc$^{-2}$.

Typically the objects have ultraviolet luminosity ~ $10^9$-$10^{10}$ times the solar luminosity, sizes ~ 1 kpc , and comoving spatial densities varying between 0.05 and 0.01 Mpc$^{-3}$ (at redshifts between 2.5 and 6). While the co-moving spatial densities are comparable to that of present-day galaxies the UV luminosity and the sizes are similar to the nearby starburst galaxies. In these galaxies the star formation occur in small concentrated regions, rather than in galaxy-sized objects. At early epochs the dynamical timescale of the galaxy-sized objects is comparable to the age of the universe, which suggests that one may be witnessing the first star formation associated with the initial "collapse of galaxies". These objects could be proto-galactic regions of star formation associated with the progenitors of the present day normal galaxies at very early epochs.

For 0000-D6, 0201-C6 and cB58, the redshift of the strong interstellar absorption lines is at z=2.96. The peak of the Lyα emission arise at a relative velocity 800 km s$^{-1}$, while the red wing extends to ~ 1500 km s$^{-1}$, and the blue wing is sharply absorbed. This P-Cygni type profile can be understood as originating in an expanding envelope around a HII region. The unabsorbed Lyα photons are back-scattered from the receding part of the nebula. The systemic velocity of the star forming region is ≈ 400 km s$^{-1}$ (as measured from the wavelength of weak photospheric lines from O-stars). The relative velocities of the interstellar , the stellar and the nebular lines point to large scale outflows. Generally less energetic outflows are seen in local starbursts.

At z ~3 the nebular emission lines which dominate the optical spectra of the star forming galaxies, are redshifted into the infrared H and K bands. The line widths, which reflect the overall kinematics of the star forming regions, can provide an indication of the masses involved. A detection of $H_\beta$ (or $H_\alpha$ at z < 2.5) would give a measure of the star formation rate, which can be compared with that from the UV continuum, and given the complexity of the UV absorption line spectra, the ratios of the familiar nebular lines may provide the way of estimating the metallicity of the galaxies (Pettini et al. 1997a).

*Interacting or merging systems*
The sample of luminous starburst galaxies gives evidence that they are interacting or merging systems. They show presence of two or more nuclei within a single distorted envelope, or presence of a second nearby galaxy with requisite proximity and relative brightness, together with bridges and long tails, that are hallmarks of tidally interacting disks.

In a few cases, hierarchical merging systems are also found. Sub-units merge into more massive systems, that take place in time scales about an order of magnitude shorter than the time-stretch of the probed cosmic epoch. In all cases the "merging" units have smaller luminosity than the other systems, but comparable surface brightness. Both parents and daughters of this possible merging scenario have comparable morphologies.

*Red objects*
Extremely red objects (ERO's), with R > 26 mag (colours of (R-K) > 5mag), once believed to be luminous star-forming galaxies at redshifts z >6, have been found to be ellipticals at z ≈ 0.8. More ERO's have been found around high-z QSO's, although they largely appear to be foreground objects. ERO's are mostly compact, but sometimes resolved. Spectroscopy of HR 10 (Hu & Ridgway 1994) indicates an evolved dusty system at modest redshift z =1.44. The z>3 galaxies from the 0000-263, 0347-383 and SSA22(2217-003) fields show that they are mostly compact galaxies, typically characterized by a bright core, less than 1.5 arcsec in diameter, often surrounded by a diffuse nebulosity with significantly lower surface brightness. The nebulosity extends to larger areas and is more irregularly distributed. Since the emission is directly proportional to the formation rate of the massive stars at the observed rest-frame in far-UV wavelengths, 90-95% of the stars which are being formed in these galaxies are concentrated in a region whose size is that of a present-day luminous galaxy. Moreover, the characteristics of the central concentration of the star formation have a size similar to a present-day spheroid. If the morphology of the massive stars is a good tracer of the overall stellar distribution, then the light distribution of the core is consistent with a dynamically relaxed structure. The central SB of the compact z>3 galaxies is consistently close to 23 mag arcsec$^{-2}$.

*Star forming galaxies and QSO's*

Co-moving ultraviolet (2800Å) luminosity density of galaxies increases by a factor of 15 from z=0 to 1. This traces the decline in the star formation rate with time.

The evolution of the space density of the quasars is similar with that of the star formation rate. There are indications that the quasar epoch may indeed be the epoch of star formation.

Wall (1997) has found a striking similarity between the trend with redshift of the space density of flat-spectrum QSO's and the evolution of star formation rate (SFR), and Dunlop finds that the evolution of the radio luminosity density of luminous radio sources is similar to the evolution of the UV luminosity density of star forming galaxies.

The shape of the quasar luminosity function changes shape with redshifts. The slope at high luminosity is significantly steeper at redshift $z \sim 2$ than it is at $z \sim 0.5$. At $z \sim 1$ the luminosity function may be described by a single power-law. At $z >1$ it can not be described by a single power-law. There is a break at $M_B \sim -26.5$. The amount of density evolution is greatest at intermediate ($M_B \sim -26$), and appears to be less at higher luminosity. At $M_B \sim -29$ the comoving space density of quasars may have remained roughly unchanged since z =2 !

There is an increase of about a factor of 10 in the SFR between the present epoch and z=1, which is consistent with QSO's (Boyle & Terlevich 1997 ). This corresponds to a redshift evolution in the mean luminosity($L^*$) of both galaxies and QSO's of the form $L^* \propto (1+z)^3$, which is also similar to the observed evolution in the infra-red luminosity of IRAS galaxies, albeit over a much lower redshift range (z<0.1). Even at redshifts higher than z=1 the agreement between QSO and galaxy samples, both in the location of the maximum and the high redshift decay rate is remarkable.

The observed luminosity function of QSO's and its redshift evolution can be explained with a starburst model for the formation of the cores of the elliptical galaxies at high redshift. The substantial fraction of the emitted luminosity in the optical/UV spectrum of QSO's is indeed associated with nuclear starburst.

*Emission and sepctra*

*Spectra*

The star forming galaxies in the nearby universe have characteristic visible spectra dominated by strong, narrow emission lines of hydrogen, oxygen and nitrogen produced in the nebulae associated with newly formed massive stars. Compared with these strong emission lines, the underlying blue continuum radiation from the stars themselves is relatively faint. However, at redshift ~ 3 the portion of the spectrum is the rest frame of the UV between 1000 and 2000Å. The strongest emission line in UV is the Lyman-alpha line of neutral hydrogen at a rest wavelength of 1215.67Å. However the Ly-alpha photons are mostly absorbed and subsequently scattered by HI atoms and very difficult to detect. The limit of the Lyman series near 912 Å (the wavelength of the photons with sufficient energy to ionize hydrogen), produces an obvious break in the far UV-spectrum of any star forming galaxy. This Lyman break has a three-fold origin: The intrinsic drop of the spectra of hot O- and B-stars which dominate the integrated spectrum at UV wavelengths, the absorption by the interstellar medium within a star forming galaxy, and the opacity of the intervening intergalactic medium. This last effect, well quantified by QSO absorption line spectra, is the overriding factor, which determine the colour of high-redshift galaxies.

Lyman-alpha break technique has been used to find high- redshift galaxies that concentrate at z=3.09. The spike may be example of today's rich clusters of galaxies caught in the evolution, when they were beginning to break away from the Hubble expansion. The Lyman-alpha objects could be progenitors of today's luminous galaxies, rather than small fragments undergoing an intense and short-lived burst of star formation. The spectra show many similarities with those of the nearby starbursts: a) the UV luminosity and implied star formation rate, b) evidence for the presence of dust and the corresponding UV extinction, and Ly-alpha emission and large scale velocity fields in the interstellar medium.

Recent observations of the normal galaxies at redshifts $z \approx 3$ have demonstrated that the Lyman-limit spectral break, and Ly$\alpha$-forest spectral decrement, which arise due to photoelectric absorption by the neutral hydrogen along the line-of-sight, constitute the most prominent spectral signatures for the very distant galaxies. The Lyman break objects in the redshift range $3.0 \leq z \leq 3.5$ represent about 1.3% of all objects to 25.0 mag, and 2.0% of all objects in the magnitude range $23.5 \leq R \leq 25.0$.

*Absorption*
At redshift z < 2.3, the spectra exhibit no significant absorption by intervening material. They are similar to the redshifted spectra of present-day galaxies. At redshift $2.5 \leq z \leq 4$ the spectra are characterized by a strong flux in the F814W and F606W images (the spectral sensitivities of these images peak at 8140 Å and 6060 Å), detectable flux in F450W images (with the sensitivity peak at 4500 Å) and no detectable flux in F300W images (with the sensitivity peak at 3000 Å). At redshift $4 \leq z \leq 5.5$ the spectra are characterized by a strong flux in the F606W images and no detectable flux in the F450W and F300W images. At redshift z > 6 the spectra are characterized by a strong flux in the F814W images and no detectable flux in the F606W, F450W and F300W images. The spectra of the objects at z > 6 are consistent with redshifted Lyman-limit absorption of high-redshift galaxies.

*X-ray emission at (0.4-10.0 keV) from M82 and NGC 253*
The X-ray spectra of the local starburst galaxies M82 and NGC 253 are very similar to the spectra of low-luminosity AGN's and LINER's.

X-ray emission is concentrated both in the nucleus and extended region along the minor axis to scales up to ~ 10 kpc. Much of the central emission is from point sources. It is likely that some of the point sources have masses > 10 $M_\odot$. The spectrum is complex with at least two components. The nuclear emission is probably consisting of the emission of unresolved supernovae and SN-heated ISM's, and the point sources which are very luminous (can be young supernovae).

The soft extended halos ( for kT< 0.5 keV) contribute negligibly in the 0.5-2.0 keV bandpass , compared to the flux originating from the region of a 6' radius centred on the nuclear source. The double Raymond-Smith model gives a fit to the M82 spectra, while the Raymond-Smith plus power-law model gives a better fit to the NGC 253 spectra. The hard and soft components are absorbed well in excess of the galactic columns ( $1.6 \times 10^{20}$ cm$^{-2}$ for M82 and $4.3 \times 10^{20}$ cm$^{-2}$ for NGC 253). The soft-component temperatures in the two cases are similar ( at 0.6 and 0.8 keV), while the hard component appears to be harder in M82 ( with kT ~ 11 keV) than in NGC 253 (with kT ~ 7keV). The temperature in the disk and nuclear regions is ~ $10^7$ K, with temperature decreasing with radius out in the halos (beyond ~ 1 kpc). In NGC 253 the soft (<2 keV) component is dominated by a disk and a nuclear thermal component with kT ~ 0.8 keV. In M82 the soft component has approximately equal contributions from ~ 1.1 keV gas in the nuclear and disk regions, and ~ 0.4 keV gas in the halo. These are consistent with the starburst driven wind. The outflow velocity of the X-ray emitting gas in M82 may be as high as 1000-3000 km s$^{-1}$, although the outflow velocity of the C gas in M82 is estimated to be less than 500 km/s.

*Infrared iron emission-line*
Iron is heavily depleted on dust in the ISM. Fast shocks propagating in the ISM (> 100 km/s), as those produced by superwinds, can destroy dust grains through a sputtering process, or grain-grain collisions, and replenish the ISM with gas-phase iron. The gaseous iron is then collisionally excited in the cooling post-shock gas, and produces the infrared emission. Supernova remnants show an enhanced FeII/H ratio up to about 1000 times more than HII regions. In galaxies, the FeII emission appears to be positionally coincident with radio emission from SNR's and with regions of star formation.

*Variability*
Variability has been observed in starburst galaxies on timescales of weeks to years. This implies that a significant amount of the emission originates from a compact (< 1pc) region. On the other hand, rapid variability on the timescales of a day, or less is not observed (rapid variability is most common in objects with 2-10 keV luminosity in the range of ~ $10^{42}$-$10^{43}$ erg s$^{-1}$, but vanishes below ~ $10^{41}$ erg s$^{-1}$). Long term variability, by a factor of up to 1.7, corresponding to a 2-10 keV flux in the range $1.4$-$2.4 \times 10^{-11}$ erg cm$^{-2}$ s$^{-1}$ ,and 2-10 keV luminosity in the range $4.2$-$7.4 \times 10^{40}$ erg$^{-1}$ , was found.

These results imply a connection of the soft component with the warm gas with kT ~ 0.6-0.8 keV, possibly from an SNR-heated ISM, and starburst-driven winds. In some cases a hard component may also be due to starburst activity, resulting from compact supernovae.

## VI.2. Galaxy Formation

*Evolution of galaxies*
Miralda-Escude & Rees (1997) and Rees & Miralda-Escude (1997) have pointed out that the physics of the galaxy formation process is different depending on whether galaxies form by accreting neutral or photoionized gas, mainly because of the cooling efficiency is different in the two cases. They have speculated that the first objects to collapse during the dark ages of the universe are likely to be on very small scales of masses ~ $10^8$ solar mass. Under these circumstances the brightest sources at the very high-redshifts are likely to be Type II supernovae.

Babul and Rees (1992) have argued that the collapse of the dwarf galaxy-scale masses is inhibited until z~ 1 due to phtoionization of the IGM by the metagalctic UV radiation. They predict that the low-mass galaxies will form at recent epochs that fade after a star formation at 0.5< z <1.0. These low luminosity, low surface brightness objects are claimed to dominate the blue number counts at faint magnitudes in redshift surveys(R. S. Ellis 1997).

Photometric redshifts for galaxies brighter than $I_{ST}$=28 in the Hubble Deep Field (Gwyn & Hartwick 1996) show two peaks: one at z ~ 0.5 and another at z ~ 2.5. The luminosity function show strong evolution: The brightest galaxies are 4 magnitudes brighter than their brightest counterparts and the faint galaxies are fewer in number. The double peaked redshift distribution and the evolution of the luminosity function can be understood if larger galaxies form stars early at z ~ 3 and if star formation is delayed in the dwarf galaxies until z ~ 1.

Interactions of large gas rich spirals may also lead to the formation of gas-rich, dark-matter poor dwarf galaxies by the fragmentation of extended, or unbound tidal tails. Such dwarfs would form in small numbers near major mergers and would be seen only in the vicinity of recently merged spirals.

The clumps from fragmentation of thin shells have low specific angular momentum and are likely to collapse into relatively compact structures, with a brief star forming phase until supernovae remove, or disperse the bulk of the remaining gas. Since the swept up material are from the IGM at high-redshift and not very enriched with metals, such dwarfs are expected to be metal poor. The first few supernovae can entirely disrupt the gas disk, truncating star formation. Therefore the number of these galaxies is likely to decline at lower redshifts, because they fade away after the initial burst of star formation. As they form in a sheet around the quasar, these dwarfs are expected to be highly clustered.

During the last few years a number of deep redshift surveys and HST imaging have studied the evolutionary state of galaxies at z ≤ 1.5. Studies show that the later and less bright types of luminosity functions have undergone a significant amount of evolution in number density and/or luminosity (of star formation rate) over the cosmic epochs, while the bright and more massive galaxies have remained substantially unchanged during the same period.

Lilly et al. (1995) have concluded that there is no evolution in either the number density or luminosity function of galaxies redder than a present Sbc galaxy over the redshift range 0 < z<1. In contrast a strong evolution of the luminosity function of galaxies bluer than a present Sbc galaxy at z>0.5 is found. A large fraction of the latter are irregular/peculiar galaxies.

The class of irregular/ merger galaxies, which are relatively rare at bright magnitudes, makes up about a half, or a third of all galaxies with $I_{AB}$ ~ 25. The median redshift at this apparent magnitude is z ~ 0.8.

The giant ellipticals and spirals at z~ 1 turn out to be mature slowly-evolving galaxies, with larger star formation than at z ~ 0. However while massive galaxies at z ~ 1 are similar to galaxies at z ~ 0, only the sub-L* population undergoes rapid evolution at z<1. There is a steady increase in the mass of actively star forming systems with an increasing redshift ( Cowie et al. 1997): Out to z ≈ 0.4 only 0.01-0.1 L* systems undergo significant star formation, while in the range 0.8 < z <1.6 the systems that are more massive ( in some case approaching L*), are observed to be in formation. Generally the local starburst systems have low masses. Moreover, the rapidly evolving population of faint blue galaxies, which dominates the less massive systems at moderate redshifts, undergo bursts of star formation at enhanced rates compared to the local starbursts.

*MgII absorbers*
The survey of MgII absorption selected galaxies by Bergeron and Boisse (1991) has shown that at z ~ 0.4 all field galaxies brighter than L>0.3L* should have extended gaseous halos with a typical radius ~ 75 $h_{50}^{-1}$ kpc. Over the redshift interval 0.2< z <1 there is no significant evolution (Steidel 1994) in the space density, or K-luminosity, or B-K color for the quasar absorption line selected galaxies.

The apparent magnitude range for z~ 0.4 absorbers is -23< $M_r$<-19.5. At this redshift the r-band roughly coincides with the redshifted B-band. At higher redshifts z~ 0.8-1.0 this corresponds to an apparent magnitude range of 25.3<$M_r$ <21.8 and $M_r$=23.7 for a L* galaxy.

The high redshift sample ( z ~ 0.7 -1.3) cover the magnitude range 21.0<$M_r$<23.1 leading to k-corrected B absolute magnitude of -23.7 <$M_{AB}$<-21.0. All the galaxies have luminosity similar, or brighter than present day L* galaxies. Their spectral types cover the whole range from elliptical to the irregular galaxies with however predominance of late type galaxies (Scd and Im).

At z~1 intense star formation activity is occurring in galaxies with a Ba+CaII break smaller than 1.55, i.e. about half of the MgII high redshift absorber subsample (Guillemin and Bergeron, 1997).

*Clustering*
In the local universe the early type galaxies (E/S0) are more strongly clustered than the late types (Sp/Irr). In the intermediate redshift similar clustering exists, with redder and more luminous systems more strongly clustered than the blue and fainter counterparts.

*UV Luminosity*
Typical high-z galaxy with R ~ 24.5 and z=3 has UV luminosity ~ 800 times greater than the brightest star cluster in the irregular galaxy NGC 4214 and exceeds by a factor of ~ 30 that of the most luminous local example, the Wolf-Rayet galaxy NGC 1741 which contains ~ $10^4$ O-type stars.

*Ly-alpha absorbers*
The comoving density of neutral hydrogen present in damped Lyman-alpha clouds peaks at z ~ 3 when it was comparable to the mass in baryon seen in galactic disks today (Storries-Lombardi et al. 1996). The decline in the abundance of the neutral hydrogen clouds seems to be accompanied by a gradual build up of their metal contents.

The metal-lines in Ly-alpha absorbers rapidly evolve with redshifts from z~ 4 to z ~1, and span a wide range of dynamical processes. At z~ 2, Ly-alpha clouds contain the majority of the baryon content of the universe. At lower redshifts, they are associated with lower surface brightness and/or dwarf galaxies, or the outer disks and halos of high surface brightness galaxies, or the remnants left over from the formation of the galaxies and/or galaxy groups that have undergone interactions.

*High-ionization metal lines*
A limited number of strong-metal-line species have been seen in high ionization transitions at z ~2.5. Relative to z~2.5 the meta-galactic UV background flux (UVB) at z ~1 is reduced by a factor of ~ 5, and its shape may be softened by stellar photons escaping bright field galaxies. Thus IGM ionization conditions have evolved so that low ionization species, especially MgII doublet and several of the stronger FeII transitions, are detectable in Ly-alpha clouds. At z~2.5, CIV absorption is detected in ~ 75% of all Ly-alpha clouds. The metalicity of z~2.5 Ly-alpha clouds is Z/Z-solar~-2 and fairly uniform.

*Ly-alpha emission*
Detected in about 75% of the galaxies and is always weaker than expected on the basis of the UV continuum luminosity. The main reason for the weakness of Ly-alpha line is resonant scattering in an outflowing interstellar medium. When detected the emission line is generally redshifted by upto several hundred km/s relative to the interstellar absorption lines. In the best observed cases its profile is asymmetric.

**Connection between quasars and galaxy formation.**

*Redshift evolution of quasars*
Over the first few Gyrs of the life of the universe the population of the bright QSO's grows and the optical luminosity functions rise. After z ~ 3 the optical luminosity functions fall to nearly the level of the local Seyfert 1 nuclei. The transitional mass from a large galaxy to a small group is around 5x $10^{12}$ $M_\odot$. In hierarchical cosmologies this corresponds to z $\cong$ 3 $\pm$ 0.5. This is where the bright QSO's peak.

The quasar formation epoch may be the epoch of massive galaxy formation due to mergers (Kats et al. 1992 & Krivitsky and Kontorovich et al. 1992). Observations of galaxy formation of blocks through merging process in the redshift range 2.6<z<3.9 allowed Clements and Couch (1996) to conclude the possibility that an epoch of the galaxy formation was discovered. The observations of sub-galactic blocks at z=2.39 (Pascarelle et al., 1995) and their relations to galaxy formation were discussed for the rich groups which were discovered in connection with faint blue galaxies.

# VII. Conclusion

*i) GRB's occur in the environment of interacting galaxies*

The observations of the counterparts of gamma-ray bursts, so far observed, indicate that the bursts occur offset from the centre of the host galaxies. In fact, GRB 970508 could have taken place outside the stellar disk of a «host» galaxy. These galaxies show elongated spiral-like features.

The associations of the gamma-ray bursts with the environments of interacting galaxies, where most red QSO's are observed, seem to be present. Extremely red objects, with R magnitude >26, believed to be star forming galaxies at high redshift, have been found around high-z QSO's. They are mostly compact systems surrounded by diffuse nebulosity, that extend to larger areas, and irregularly distributed. The stars which are being formed in these galaxies are concentrated in a region whose is comparable to that of the present day luminous galaxies.

*ii) GRB's occur in the star-forming regions*
The host galaxy of the GRB 970228 has been found to possess a R-magnitude ~25. The optical luminosity of the point source, that represents the burst location, decays following a $t^{-1}$ power-law. For GRB 970508, a magnitude R ~ 26 has been suggested for the possible underlying galaxy. The decline of the brightness in this case can be fitted with a $t^{-1.14}$ power-law. The R-band magnitude of the host of the GRB 971214 is predicted to be 25.6 mag, while its decay index is -1.38, which is similar to the data for GRB 970228 in the first days. Moreover, the magnitude of the OT host galaxy for GRB 980326 and GRB 980519 are believed to be R=25.5± 0.5 and R ~ 24 or higher.

The apparent R-magnitude of a galaxy having a luminosity of a present day normal galaxy at redshift of ~ 0.8, will be ~ 24. Moreover, the redshift sample of the galaxies in the range z ~ 0.7 - 1.3 with luminosity brighter, or similar to the present day L* galaxies, cover the whole range of the galactic morphology, though the late type galaxies (Scd and Im) seem to be predominant. Luminous galaxies placed at redshift z ~3, or dwarfs around z ~ 1, will show the magnitude range as has been observed for the counterparts. We conclude that these host galaxies lie in the star-forming regions at high redshifts and appear in the environment where quasars are found.

*iii) GRB's associated with quasars and BL Lacs*
The observation of the featureless spectrum, except for a few absorption lines, in GRB 970508, and the detection of the emission and absorption features in GRB 971214 after the initial burst, indicate that the bursts could be associated with phenomena related to BL Lacs, that are found among the reddest of the red quasars. Some of the MgII absorbers in BL Lacs may be intrinsic, while there is definitely a correlation between the absorbing gas and the featureless spectrum. This leads us to infer that the MgII systems, detected in GRB 970508, and the Ly-α absorbing gas, found in GRB 971214, could indicate environments of interacting galaxies in star-forming regions at high-redshift as they are in Quasars and BL Lacs.

*iv) Jets cause the evolution from high-energy to low-energy emission features*
The radio data from GRB 970508 demand a scenario where shock induced plasma emission takes place, which is similar to intra-day variable QSO's. The changes in the scintillation pattern of the radio emission also indicate that the radio emission takes place from a relativistic plasma. In GRB 980425 the radio flux increased dramatically after about a week from the time of the burst. This emission can arise due to relativistic shocks. A core-jet structure of an expanding plasma, that emit synchrotron emission , starts out opaque and becomes transparent at lower frequencies, can explain such emission pattern. The steep spectra observed in the GRB 980228 could arise from jets with low-surface brightness. In the case of QSO 1156+295, near the peak of the optical outburst, radio variability over a few days was found, as for GRB 980425. There was a large radio flare in QSO 1156+295, that began a few months later.

These observations may indicate that the evolution of the emissions starting from the high-energy gamma-rays to the low-frequency radio band, as well as the long-lasting X-ray afterglow could be caused by plasma, that are ejected in an explosion. When the relativistically moving plasma slows down, and the sub-relativistic plasma catch it up from behind, long-lasting afterglow are produced as in the fireball models.

*v) Supernova*
Although the supernovae decay at slower rate in the visible wavelengths, than what are observed with the optical counterparts, they can decay much faster in UV. Thus a supernova at high-redshift could appear to decay in the visual, which is UV in the rest frame, at the observed rate. The fast decline of the counterparts indicate that they could be supernovae decaying in UV at high frequencies.

Our final conclusion is that the GRB's are generated by very massive stars, that end up their life as «hypernovae», in compact star forming regions, that harbour around large galaxies undergoing interactions. The gamma-ray light curves can be explained by invoking inverse Compton scattering of soft photons from blobs moving in jets ejected by such stellar explosions. Relativistically moving shells as well as jets can possibly explain the characteristics observed for the afterglow.

In the light of this study it will be necessary to investigate the mechanisms by which massive stars may collapse and give birth to relativistic jets before one may resolve the mystery of the gamma-ray bursts. Such investigations may throw new perspective and theories regarding star formation.

**Acknowledgment**
*I should like to express my gratitude to Professor Hans Kolbenstvedt and Professor Erlend Østgaard for extending a cordial and helpful relation with the Department of Physics, NTNU(Lade). I should especially thank Professor Østgaard for bringing the issue of the gamma-ray bursts to my attention. I deeply appreciate the help that Professor Razi Naqvi has extended to me lately.*

# References

*Babul, A. and Rees, M. J., 1992, Month. Notice of the Roy. Astron. Soc. 255, 346*
*Bahcall, J. N., Kirhakos, S., Saxe, D.H., Schneider, D. P., 1997, Astrophys. Journ. 479, 642*
*Bergeron, J. and Boisse, P., 1991, Astron. and Astrophys. 243, 344*
*Bond, H. E., 1997, IAU Circ. 6654*
*Bowen, D., Baldes, C., Pettini, M. , 1996, Astrophys. Journ. 472, L77*
*Boyle , B. J. and Terlevich, R., 1997, preprint:astro-ph/9710134*
*Brainerd, J. J. , 1992, Astrophys. Journ. 394, L33-36*
*Bremer, M. et al. 1998, Astron. Astrophys. 332, L13*
*Briggs, M. et al. 1998, IAU Circ. 6856*
*Burbridge, E. M. et al. 1996, Astron. Journ. 112, 2533*
*Caraveo, P. A. et al. 1997, preprint:astro-ph/9707163*
*Carilli, C. L. et al, 1997, preprint, astro-ph/9709030*
*Carilli, C. L. et al., 1998, preprint:astro-ph/9801157*
*Castander, F. J. et al. 1997, IAU Circ. 6791*
*Celidonio, G. et al. 1998,IAU Circ. 6851*
*Churchill, C. W. , 1998, Astrophys. Journ. 499, 677*
*Churchill, C. W. , Steidel, C. C., and Vogt, S. S., 1996, Astrophys. Journ. 471, 164*
*Clements, D. L. and Couch, W. J., 1996, preprint: astro-ph/9605130*
*Cohen, R. D. et. al. 1987, Astrophys. Journ. 318, 577*
*Costa, E. et al. 1997a, IAU circ. 6572*
*Costa, E. et al. 1997b, IAU circ. 6576*
*Costa, E. et al. 1997c, IAU Circ. 6649*
*Cowie, L. L. et al. 1997, preprint: astro-ph/9702235*
*de Vries , W. H. et al. 1997, Astrophys. Journ. Suppl. 110, 191*
*Dermer, C. D. et al. 1992, Astron. Astrophys. 256, L27*
*Dermer, C. D. and Schlickeiser,R., 1994, Astrophys. Journal. Suppl. 90, 945*
*Diercks, A. H. et al. 1998, Astrophy. Journ. 503, L105*
*Djorgovski, S. G. et al. 1997a, IAU Circ. 6660*
*Djorgovski, G. et al. 1998, GCN message No. 41*
*Djorgovski, S. et al. 1997b, IAU Circ. 6655*
*Elias, J. H. et al. 1978, Astrophys. Journ. 220, 25*
*Ellis, R. S., 1997, Int. Astron. Union Symp. 186, 2*
*Filippenko, A. V. , in Supernovae, The Tenth Santa Cruz Summer Workshop in Astronomy and Astrophysics, ed. by Woosley, S. E., Springer Verlag, 1990, 467*
*Fisher, K. B., Bahcall, J. N., Kirhakos, S. and Schneider, D. P. , 1996, Astrophys. Journ.. 468, 469*
*Frail, D. A. et al. 1997, IAU circ. 6662*
*Frail, D. A. et al. 1997, Nature 389, 261*
*Gal, R. R. et al. 1997, American Astron. Soc. Meeting 191*
*Galama, T. J. et al. 1997, Nature 387, 479*
*Galama, T. J. et al. 1998a, IAU Circ. 6895*
*Galama, T. J. et al. 1998b, IAU Circ. Nos. 6895 & 6898*
*Groot, P. J. et al. 1997, IAU circ. 6584*
*Grossman, B. et al. 1998, GCN message No. 35*
*Guarnieri, A. et al. 1997, Astron. Astrophys. 328, L13*
*Guillemin , P. and Bergeron, J. , 1997, Astron. Astrophys. 328, 499*
*Gwyn, S. D. J. and Hartwick, F. D. A. , 1996, Astrophy. Journ. 468, L77*
*Halpern, J. et al. 1997, IAU Circ. 6788*
*Heiese, J. et al. 1997a, IAU Circ. 6654*
*Heiese, J. et al. 1997b, IAU Circ. 6787*
*Hu, E. M. and Ridgway, S. E., 1994, Astronom. Journ. 107, 1303*
*Hunter, S. D. et al. 1993, Astrophy. Journ. 409, 134*
*Jager, R. et al. 1997, Astron. Astrophys.*
*Jaunsen et al. 1998, GCN message No. 78*
*Jones, D. L. et. al. 1984, Astrophys. Journ., 284, 60*
*Kats, A. V. et al. 1992, Sov. Astr. Lett. 17, 96*
*Kippen, R. M. et al. 1998, GCN message No. 87*
*Kouveliotou, C. et al. 1997, IAU Circ. 6660*
*Krivitsky, D. S. And Kontorovich, V. M., 1998, preprint: astro-ph/9801195*
*Krolik, J. H. & Pier,E. A., 1991, Astrophys. Journ . 373, 277*
*Kulkarni, S. R.. et al. 1998, Nature 393, 35*

*La Franca, F. et al. 1998, preprint: astro-ph/9802281*
*Landau , R. et al. 1983, Astrophys. Journ. 268, 68*
*Lange, N., in Supernovae, The Tenth Santa Cruz Summer Workshop in Astronomy and Astrophysics, ed. by Woosley, S. E., Springer Verlag, 1990*
*Lilly, S. J., Le Fevre, O. et al. 1996, Astrophys. Journ. 460, L1*
*Camenzind M., Krockenberger,M., 1992, Astron. Astrophys. 255, 59*
*Madejski , G. et al. 1996, preprint: X-Ray-95-18 (CERN: SCAN-9604073)*
*Madejski, G. et al. 1996, Astrophys. Journ. 459, 156*
*Marshall, F. and Takeshima, T., 1998, GNC message No. 58*
*Mc Breen et al.,1993, Astron. Astrophys. Suppl. 97, 81*
*Metzger, M. R. et al. 1997a, IAU Circ. 6655*
*Metzger, M. R. et al. 1997b, IAU Circ. 6672*
*Metzger, M. R. et al. 1997c, Nature 287, 878*
*Meurer, G. et al. 1997, Astron. Journ. 114, 54*
*Miralda-Escude & Rees, M. J. , 1997, preprint: astro-ph/9707193*
*Nilsson, K. et al. 1996, Astron. Astrophys., 314, 754*
*O'Dell, S. L. et. al. 1988, Astrophy. Journ., 326, 668*
*Pascarelle, S. M.  et al. 1995, preprint: astro-ph/9512033*
*Pettini, M. et al. 1997a, Astrophys. Journ. 478, 536*
*Pettini, M. et al. 1997b, preprint: astro-ph/9707200*
*Pian, E. et al. 1997, preprint: astro-ph/9712082*
*Piro, L. et al. 1997a, preprint: astro-ph/9710355*
*Piro, L.  et al. 1997b, IAU Circ. 6656*
*Pooley, G. Et al. IAU Circ. 6670*
*Rees, M. J. and Miralda-Escude, 1997*
*Rej, A. and Østgaard ,E., 1997, TPS, NTNU, No.17 (astro-ph/9807113)*
*Rhoads, J. , 1997, IAU Circ. 6793*
*Rieke , G. H. et al. 1976, Nature 260, 754*
*Roland, J.  et al. 1994, Astron. Astrophys. 290, 364*
*Sahu, K. C. et al. 1997, preprint: astro-ph/9705184v2*
*Shaviv, N. J.  & Dar, A.., 1995 , Astrophys. Journ. 447, 863*
*Smith, I. A. , Gruendl, R. A. et al. 1997, Astrophys. Journ. 487, L5*
*Soffitta, P. et al. 1998, IAU Circ. 6884*
*Standke, K. et al. 1996, preprint:astro-ph/9512135*
*Steidel, C., Dickinson, M., 1994, preprint:astro-ph/9409061*
*Stickel , M. et al. 1991, Astrophys. Journ. 374, 431*
*Stocke, J. T. & Rector, T. A. , 1997, preprint: astro-ph/9709111*
*Stocke, J. T. et al. 1991, Astrophys Journ. Suppl. 76, 813*
*Storries-Lombardi, L. J. et al. 1996, Month. Notice Roy. Astron. Soc. 283, L 79*
*Taylor, G. B.,  1997, Nature 389, 263*
*Van den Bergh, S., 1992, Astroph. Journal. 390, 133*
*Van Paradijs, J. et al. 1997, Nature 386, 686*
*von Montigny, C.  et al 1995., Astrophys. Journ. 440, 525*
*Wall, J., 1997, RGO preprint 268*
*Wang, L. and Wheeler, J. C. , 1998, preprint: astro-ph/9806212*
*Webb, J. R. and Smith, A. G., 1989, Astron.  Astrophys. 220, 65*
*Yamada, T. et al., 1997, preprint, astro-ph/9707197*